\documentclass[a4paper,10pt,aps,prb,twocolumn,superscriptaddress]{revtex4-2}

\usepackage[utf8]{inputenc}
\usepackage{mathtools}
\usepackage{amsmath, amsthm, amssymb, amsfonts}
\usepackage{hyperref}
\usepackage{graphicx}
\usepackage{xcolor}
\usepackage{verbatim}
\usepackage{comment}

\begin{document}


\title{Strain-Controlled Magnetic Phase Transitions through Anisotropic Exchange Interactions: A Combined DFT and Monte Carlo Study}

\author{Sudip Mandal}
\affiliation{Theory Division, Saha Institute of Nuclear Physics, A CI of Homi Bhabha National Institute, Kolkata 700064, India}

\author{Mihir Ranjan Sahoo}
\affiliation{Graz University of Technology, 8010 Graz, Austria}

\author{Kalpataru Pradhan}
\affiliation{Theory Division, Saha Institute of Nuclear Physics, A CI of Homi Bhabha National Institute, Kolkata 700064, India}

\date{\today}

\begin{abstract}
Epitaxial strain provides a powerful, non-chemical route to tune the properties of
functional materials by manipulating the coupling between spin, charge, and lattice
degrees of freedom. Using density functional theory (DFT) calculations and $\rm BiFeO_3$
as a model system, we first demonstrate how epitaxial strain exactly leads to
anisotropic magnetic interactions where the exchange coupling along the $c$-axis
differs from that in the $ab$-plane. We show that subtle structural modifications, 
specifically the distortion from a cubic to a tetragonal lattice, drive a 
magnetic phase transition from a G-type to a C-type antiferromagnetic (AF) phase.
The anisotropy in magnetic interactions, which becomes prominent in the lower
symmetry tetragonal phase, provides a direct link between the structural distortion
and the potential change in magnetic ordering. For a more comprehensive
study, we next investigate the role of strain in driving magnetic phase transitions
within a half-filled one-band Hubbard model in three dimensions. In this framework,
strain is introduced through anisotropic hopping processes between nearest- and
next-nearest-neighbor sites, inspired by the DFT calculations. Using a semiclassical
Monte Carlo (s-MC) approach, we construct ground state phase diagrams in the
nonperturbative regime, which show how uniaxial strain stabilizes distinct magnetic
ground states: Compressive strain drives a transition from a G-type to a C-type AF
insulator, whereas tensile strain suppresses the C-type AF order, favoring an A-type
AF phase. Overall, our combined DFT and s-MC calculations highlight that strain is
a powerful tuning parameter for controlling competing magnetic phases by governing
exchange coupling mechanisms in correlated systems, offering valuable insights
for the design of strain-controlled materials.
\end{abstract}



\maketitle

\section{Introduction}

Competing spin interactions are central to the unusual properties of correlated
materials in physics~\cite{Dagotto, Dagotto2, Avella, Liu}. 
The interplay between these interactions often leads to novel and intriguing 
phenomena~\cite{Sachdev, Lacroix, Sachdev2}. Interestingly, when the strengths of 
these interactions are nearly balanced, the system becomes extremely sensitive to
small perturbations like temperature, pressure, or strain, resulting in complex,
rich phase diagrams where minor changes can trigger transitions between different 
magnetic phases~\cite{Lee, Tsymbal}. This sensitivity is mostly due to the strong
correspondence that exists between spin interactions and the crystal lattice. 
Since magnetic properties are intrinsically linked to the underlying structure,
lattice distortions significantly influence magnetic behavior, directly altering 
exchange interactions and spin configurations~\cite{Hanbucken, Zhong}. Therefore,
alterations to the crystal structure can directly impact the magnetic ground state. 
Understanding these coupled structural and magnetic transitions is crucial for 
advancing condensed matter physics and exploring applications in new technologies.

Materials exhibiting magneto-structural transitions, in which magnetic and lattice
degrees of freedom are strongly coupled, are of considerable interest from both
fundamental and applied viewpoints~\cite{Dutta, Roy}. On the fundamental side,
such transitions offer a well-defined framework to investigate how lattice symmetry
breaking influences the magnetic ordering, electronic correlations, and exchange
interactions, thereby providing valuable insights into the physics of correlated
electron systems. From an applied perspective, this strong coupling enables external
control of the magnetic states through strain, pressure, or temperature. This
tunability is particularly promising for spintronic and memory applications, where
switching between distinct magnetic states via structural modulation can enable
energy-efficient, non-volatile device operation~\cite{Yan}. A detailed understanding
of these coupled magnetic and structural transitions is therefore crucial for
advancing both fundamental condensed matter physics and functional material design.

The close connection between the lattice structure and the magnetism is clearly 
apparent in perovskite oxides, where chemical substitution typically causes 
simultaneous structural and magnetic transformations~\cite{Imada, Tokura, Dagotto3}. 
When a few of the ions are replaced by others of differing size or charge, the
resulting chemical pressure introduces internal strain, distorting the crystal
structure to accommodate the modified bonding~\cite{Zhou, Bartel}. These lattice
distortions consequently alter inter-atomic distances and bond angles, directly
tuning the superexchange pathways that govern magnetic ordering. A key example
is $\rm Sr_{1-x}Ba_{x}MnO_3$, where substituting Ba for Sr causes a progressive
tetragonal distortion that strongly impacts the magnetic ground state~\cite{Somaily, Chen}.
At low Ba content, the nearly cubic phase exhibits a G-type antiferromagnetic (AF)
order. However, as the Ba concentration rises, the increasing tetragonal distortion
drives the system away from pure G-type AF order, initially resulting in a mixed
antiferromagnetic-ferromagnetic phase before finally stabilizing a pure ferromagnetic
(FM) order. Another significant illustration is $\rm La_{0.2}Sr_{0.8}MnO_3$~\cite{Bindu, Pinsard, Chaluvadi, Sharma}. 
This material has a nominally cubic perovskite structure and is paramagnetic (PM)
above room temperature, however local $\rm MnO_6$ octahedral distortions are
already present. Upon cooling, it undergoes a dramatic magneto-structural
transition into a tetragonally distorted phase, which is intimately associated
with the emergence of C-type AF ordering.

The same underlying principle---that lattice distortions induced by ionic-size
mismatch can dictate magnetic ordering---is also evident in other Mn-based
perovskites such as $\rm CaMnO_3$ and $\rm CdMnO_3$. Differences in the ionic
radii of Ca and Cd lead to distinct lattice symmetries and, consequently, different
magnetic ground states: $\rm CaMnO_3$ crystallizes in an orthorhombic perovskite
structure and exhibits G-type AF ordering with $T_N \approx 123$~K~\cite{Zeng, Kawazoe},
whereas $\rm CdMnO_3$, which incorporates the smaller Cd ion, is stabilized
in a tetragonal perovskite lattice and displays a C-type AF ground state with
a reduced $T_N \approx 86$~K~\cite{Xu}.

Electronic instabilities, such as the Jahn--Teller effect, provide another route
to structural distortion~\cite{Bersuker, Koppel, Streltsov}. This effect, originating
from the lifting of orbital degeneracies in transition metal ions with partially
filled $d$ orbitals, drives spontaneous lattice distortions that lower the
electronic energy. In perovskites, these distortions typically manifest as elongations
or compressions of the metal--oxygen octahedra, directly coupling orbital, lattice,
and spin degrees of freedom. A notable example is $\rm LaMnO_3$, where cooperative
Jahn--Teller distortions of $\rm Mn^{3+}$ ions stabilize an orthorhombic structure
at room temperature, strongly influencing its A-type AF order~\cite{Hotta}. The
principle of electronic instability driving structural change is also evident in
materials like $\rm LaCoO_3$, which undergoes a phase transition from a rhombohedral
structure to a cubic lattice at temperatures above approximately 1200~K, a transition
often linked to spin-state changes~\cite{Kobayashi}.

While structural changes are naturally induced by intrinsic factors like ionic-size
mismatch and electronic instabilities (e.g., the Jahn--Teller effect), epitaxial strain
is one of the most commonly used extrinsic tools to modify and control material
properties by driving structural phase transitions, modifying electronic bandwidth, or
affecting magnetic ordering~\cite{Carman, Bhattacharya, Bandyopadhyay, Qi, Bhattacharya2, Dhole, Sarkar}. 
Specifically, when a material is grown epitaxially on a substrate with a
different lattice constant, the resulting lattice mismatch generates strain that
distorts the unit cell and perturbs the orbital overlap and exchange interactions~\cite{Vailionis, Mirjolet, Li}. 
Hence, resulting strain can compress or elongate bond lengths and change bond angles,
thereby tuning the strength and geometry of exchange interactions~\cite{Guo, Paris, Wakabayashi, Lupo, Mo, Paudel, Lupascu, Seo}. 
This directly affects key electronic parameters such as the hopping amplitudes and 
superexchange pathways, which are essential in determining the magnetic ground
state~\cite{Banerjee, Tsvetkov}.
    
The power of epitaxial strain in modifying bond lengths and angles---and thus tuning
electronic and magnetic behavior---is demonstrated across various material systems.
An example of strain-controlled magnetism is found in the iridate $\rm Sr_{2}IrO_{4}$.
While the bulk material shows a canted G-type AF order with N\'eel temperature
$T_N \approx 240$~K~\cite{Kim}, tensile strain enhances $T_N$, whereas compressive
strain suppresses it~\cite{Nichols, Lupascu, Geprags}. This modulation of $T_N$ is a
direct result of strain-induced changes in the Ir--O--Ir bond angles. Similarly,
$\rm EuTiO_3$, which in bulk crystallizes in a cubic perovskite structure and exhibits
G-type AF ordering at low temperatures with $T_N \approx 5.6$~K~\cite{Scagnoli, Midya, Laguta},
provides a demonstration of strain-driven transition. Under epitaxial strain, it
undergoes a cubic-to-tetragonal distortion that weakens the AF superexchange and
stabilizes a ferromagnetic order~\cite{Lin}.

$\rm BiFeO_3$ is another extensively studied example of a strain-induced magnetic
transition; in this multiferroic material, strain can drive a transition between
G-type and C-type AF orders~\cite{Fan, MacDougall, Dixit, Yang, Goswami}. In its pseudocubic
structure, the dominant nearest-neighbor (NN) superexchange interactions are nearly
isotropic, favoring G-type AF ordering. However, when strain lowers the crystal
symmetry to tetragonal, the emergence of a distinct $c$ axis introduces anisotropic
exchange interactions. These anisotropies, together with the contribution of
next-nearest-neighbor (NNN) couplings, can stabilize a C-type AF configuration. 
First-principles density functional theory (DFT) calculations have been
instrumental in clarifying these mechanisms, providing insights into the electronic
structure, total energies, and changes in magnetic exchange interactions across
different structural phases. DFT studies on $\rm BiFeO_3$ have explicitly
demonstrated the strain-induced transition between C-type and G-type AF ordering,
as well as its impact on band dispersion and optical properties~\cite{Dieguez, Ding, Escorihuela, Walden, Walden2, Tong}. 
Similar strain-controlled transitions have been reported in related perovskites, 
including $\rm BiCoO_3$~\cite{Dieguez, Escorihuela} and $\rm BiCrO_3$~\cite{Walden, Walden2}, 
underscoring the broader role of strain in tuning magnetic phases. 
Even more intricate strain-driven phase behavior has been predicted in double
perovskites: for instance, $\rm Ca_2FeOsO_6$ transitions from a G-type ferrimagnetic
phase to a C-type AF phase under compression, and further to an E-type AF
phase under tensile strain~\cite{Rout}, while $\rm Sr_2FeOsO_6$ undergoes
a C-type AF to G-type ferrimagnetic transition under epitaxial strain~\cite{Rout2}.

Such strain-induced symmetry lowering has profound consequences for both
electronic and magnetic properties. It lifts degeneracies in the electronic
band structure, alters the Fermi surface, and significantly influences transport
behavior~\cite{Dhole, Andrei}. Simultaneously, lattice distortions modify
inter-atomic distances and bond angles---particularly those involving magnetic
and non-magnetic ions such as oxygen---thereby tuning the superexchange interactions.
Thus, these structural changes directly affect the strength and geometry of magnetic
exchange, including both NN and NNN couplings~\cite{Ding, Escorihuela}.
This control over symmetry and exchange interactions is highly versatile: the
specific structural transition (like a cubic-to-tetragonal distortion) can
itself be triggered not only by epitaxial strain but also by other external
or internal factors such as temperature, magnetic field, or chemical doping.

We aim to qualitatively understand the mechanisms that drive strain-induced
magnetic phase transitions in correlated materials. To achieve this, we utilize
the half-filled one-band Hubbard model~\cite{Gutzwiller, Kanamori, Hubbard, Staudt, Arovas, Qin}
on a simple cubic lattice. Strain is modeled by introducing anisotropic
hopping amplitudes---specifically, nearest-neighbor ($t_x$, $t_y$, $t_z$) and
next-nearest-neighbor ($t'_{xy}$, $t'_{yz}$, $t'_{xz}$) terms. These directionally
dependent hopping parameters essentially modify the electronic structure by
modulating the effective bandwidth, shifting the Fermi level, and introducing
magnetic frustration. Collectively, these changes destabilize conventional
magnetic ordering, thereby paving the way for different kind of magnetic phases.

To explore the resulting electronic and magnetic properties, we systematically
vary these hopping parameters and employ the semiclassical Monte Carlo (s-MC)
technique~\cite{Mukherjee, Jana, Chakraborty, Bidika, Halder, Mandal, Mandal2}
to map the evolution of the magnetic ground states. Our focus is on the
non-perturbative regime, where the on-site repulsive Hubbard interaction ($U$) is
comparable to the non-interacting electronic bandwidth. This regime is highly
relevant for correlated materials, as it highlights the fundamental competition
between electron mobility and Coulomb repulsion---the key interplay that dictates
the emergence of new magnetic phases and transitions in such systems.

This paper is organized to first introduce the theoretical framework before
discussing the results. We begin in Sec.~\ref{sec_dft} by presenting the DFT
calculations used to estimate direction-dependent exchange couplings and establish
the realistic model parameters for $\rm BiFeO_3$ under epitaxial strain. 
In Sec.~\ref{sec_mm}, we define the anisotropic one-band Hubbard Hamiltonian 
and outline the s-MC framework used to study strain-driven magnetic and 
transport phenomena. Details regarding the derivation of the effective
Hamiltonian and the numerical procedures for computing magnetotransport observables 
are provided in Appendix~\ref{derivation_Heff} and Appendix~\ref{obs}, respectively. 
The results are then analyzed systematically: in Sec.~\ref{sec_compressive_strain}, 
we examine the effect of compressive strain on magnetic ordering, 
while Sec.~\ref{sec_tensile_strain} extends the analysis to the tensile strain
regime and reveals the associated reorganization of competing AF phases. 
Finally, the key results are summarized in Sec.~\ref{sec_con}.


\section{Estimation of hopping parameters using DFT} \label{sec_dft}

The magnetic transition, for example from G-type AF to C-type AF in $\rm BiFeO_3$,
induced by strain, likely originates from a change in the dominant magnetic exchange
pathways as the crystal structure evolves. In a cubic lattice, if AF interactions are
the only dominant interactions, they are typically isotropic, favoring the G-type AF
ordering where spins on neighboring sites are antiparallel in all three dimensions.
However, when the crystal structure transitions to tetragonal under strain, the symmetry
is lowered, introducing a unique $c$-axis different in length from the $a$ and $b$ axes.
This structural distortion leads to anisotropic exchange interactions, where the magnetic
coupling strength becomes direction-dependent. These anisotropic interactions, possibly
supplemented by the NNN exchange interactions, can then favor a C-type AF order in the
tetragonal structure. Overall, the observed magnetic transitions are attributed to
a complex interplay of factors, specifically the anisotropic exchange interactions
and the competition between magnetic exchange pathways depending on the magnitude
of the tetragonal distortion.

\begin{figure}[t!]
\centering
\includegraphics[width=0.48\textwidth]{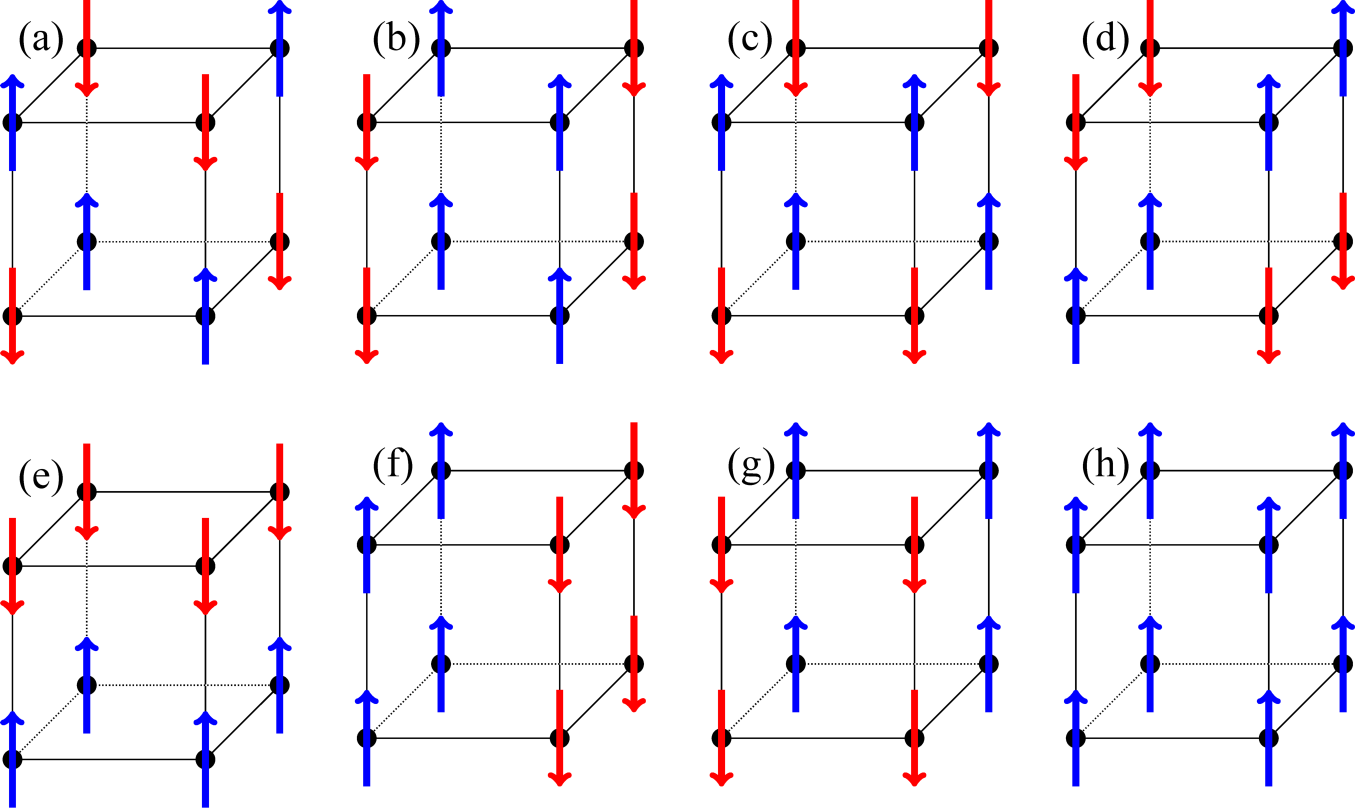}
\caption{Representative magnetic spin configurations on a simple cubic lattice,
characterized by distinct ordering vectors $\mathbf{q}$. 
(a) G-type AF: spins alternate along all three spatial directions, $\mathbf{q} = (\pi, \pi, \pi)$. 
(b) C-type AF$(\pi, \pi, 0)$: AF order in the $xy$-planes with ferromagnetic stacking
along $z$, $\mathbf{q} = (\pi, \pi, 0)$. 
(c) C-type AF$(\pi, 0, \pi)$: AF alignment in $xz$-planes with ferromagnetic coupling
along $y$, $\mathbf{q} = (\pi, 0, \pi)$. 
(d) C-type AF$(0, \pi, \pi)$: AF alignment in $yz$-planes with ferromagnetic stacking
along $x$, $\mathbf{q} = (0, \pi, \pi)$. 
(e) A-type AF$(0, 0, \pi)$: spins align ferromagnetically within each $xy$-plane,
with neighboring planes stacked antiferromagnetically along the $z$-axis; $\mathbf{q} = (0, 0, \pi)$.  
(f) A-type AF$(0, \pi, 0)$: ferromagnetic alignment in the $xz$-plane, with alternating
spin orientation along the $y$-axis; $\mathbf{q} = (0, \pi, 0)$.  
(g) A-type AF$(\pi, 0, 0)$: ferromagnetic layers lie in the $yz$-plane, coupled
antiferromagnetically along the $x$-direction; $\mathbf{q} = (\pi, 0, 0)$.
(h) FM: uniform spin alignment across the lattice, $\mathbf{q} = (0, 0, 0)$. 
}
\label{fig01}
\end{figure}

We begin by using DFT to extract realistic parameters for building a minimal model
for studying the magnetic transitions in correlated materials under epitaxial strain.
We use $\rm BiFeO_3$ as a model system for our DFT calculations. Our goal is to
connect the microscopic exchange couplings and eventually the hopping amplitudes to
the strain-induced structural anisotropy, and thereby to identify the key parameters
that control the competition between different magnetic orders. Please note that under
epitaxial strain, the lattice parameters may evolve anisotropically, providing a
route to tune the magnetic ground state. Representative magnetic spin configurations
characterized by distinct ordering vectors $\mathbf{q}$ are shown in Fig.~\ref{fig01}.

\begin{figure}[t!]
\centering
\includegraphics[width=0.28\textwidth]{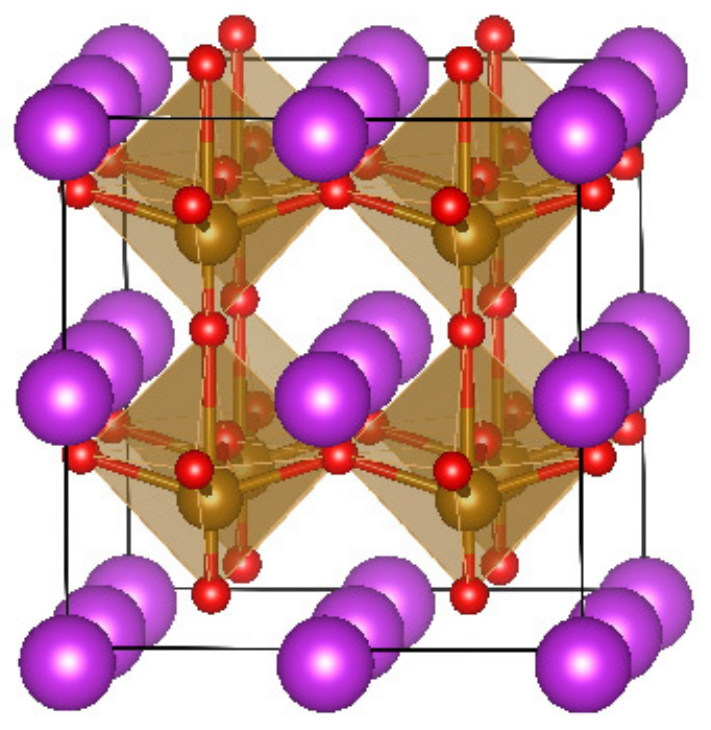}
\caption{Crystal structure of tetragonal $\rm BiFeO_3$ shown in a $2 \times 2 \times 2$
supercell. Purple spheres represent Bi ions, brown spheres denote Fe ions located
at the centers of corner-sharing $\rm FeO_6$ octahedra, and red spheres correspond
to oxygen ions. 
}
\label{fig02}
\end{figure}

All structural and magnetic properties of tetragonal bulk $\rm BiFeO_3$ are investigated
using DFT as implemented in the Vienna \textit{ab initio} Simulation Package (VASP)
code~\cite{Kresse, Kresse2}. To explore a broad range of magnetic configurations,
a $2 \times 2 \times 2$ supercell (see Fig.~\ref{fig02}) is chosen for all magnetic
calculations~\cite{Dieguez2, Escorihuela, Goswami}. A kinetic energy cutoff of $500$~eV 
is employed for the plane-wave basis set expansion. The Perdew--Burke--Ernzerhof
(PBE) version~\cite{Perdew} of the generalized gradient approximation (GGA) is adopted
as the exchange-correlation functional in conjunction with the projector augmented
wave (PAW) method. A $5 \times 5 \times 5$ Monkhorst-Pack $k$-point grid is used to
sample the first Brillouin zone for the self-consistent calculations. When the $c/a$
ratio (the ratio of the out-of-plane to the in-plane lattice parameter) is $\geq 1.1$,
the $k$-point mesh is adjusted to $5 \times 5 \times 4$ to ensure appropriate sampling
along the elongated crystallographic direction. To account for the strong on-site
Coulomb interactions associated with the localized Fe $3d$ states, we use the GGA+$U$
approach formulated by Dudarev \textit{et al}~\cite{Dudarev}. In this scheme, the
effective Hubbard parameter $U_{eff}=U-J$, is set to $4.6$~eV for Fe atoms
only~\cite{Escorihuela, Goswami}, ensuring a more accurate description of electronic correlations
in this transition-metal oxide system.

\begin{figure}[t]
\centering
\includegraphics[width=0.48\textwidth]{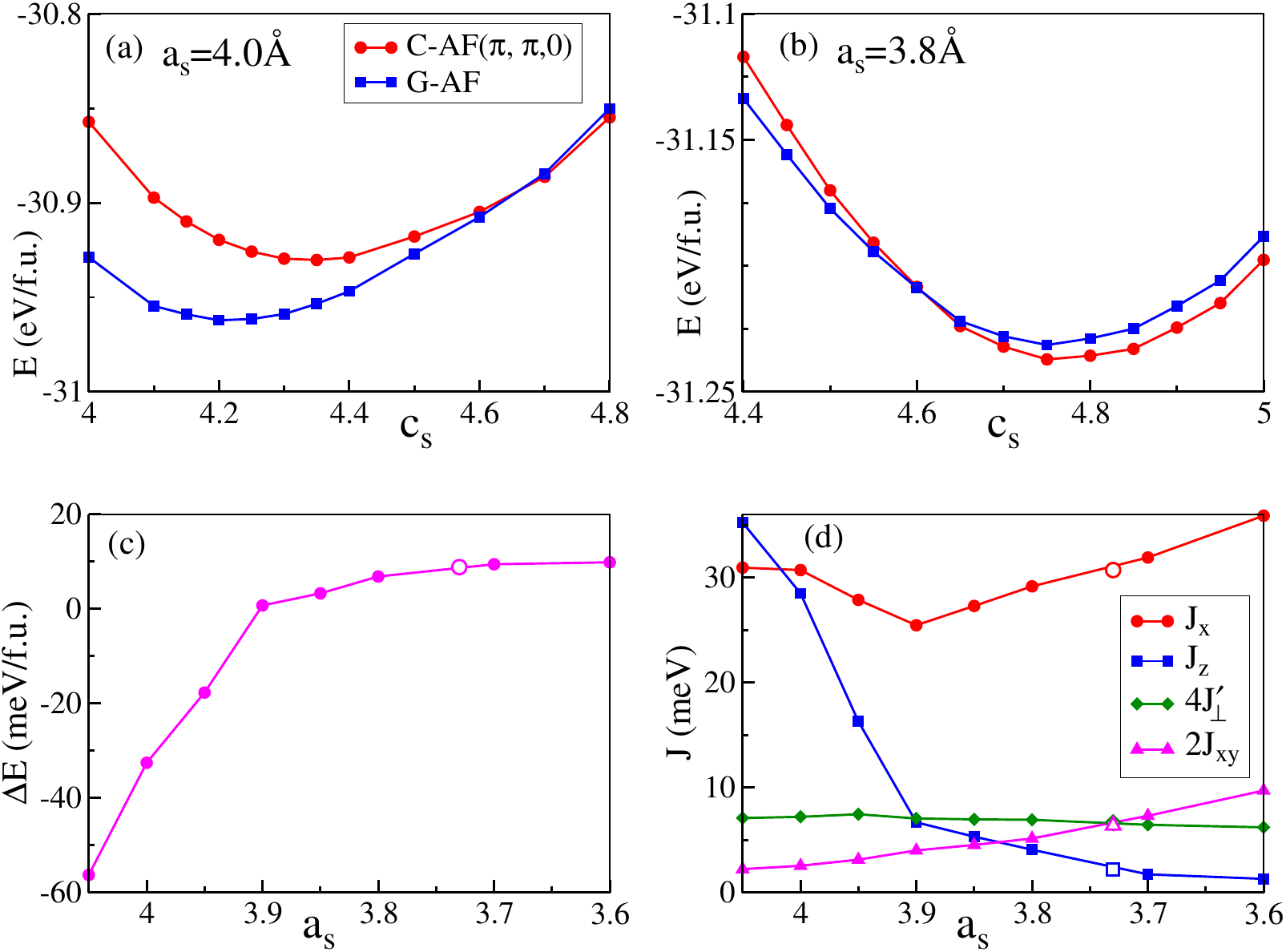}
\caption{Energies of G-type AF and C-type AF($\pi,\pi,0$) phases in $\rm BiFeO_3$ 
as a function of the out-of-plane lattice constant $c_s$, for two fixed in-plane
lattice constants: (a) $a_s = 4.0$~\AA\ and (b) $a_s = 3.8$~\AA. As $a_s$ decreases
from $4.0$~\AA\ to $3.8$~\AA, the magnetic ground state switches from G-type AF to
C-type AF($\pi,\pi,0$). (c) Energy difference $\Delta E = E_G - E_C$ as a
function of the in-plane lattice constant $a_s$, where the out-of-plane lattice
parameter $c_o$ is fixed at its optimized value, which minimizes the total energy.
As $a_s$ decreases (increasing compressive strain), $\Delta E$ gradually increases
and eventually becomes positive, indicating a transition from the G-type AF state
to the C-type AF($\pi,\pi,0$) state.
(d) Exchange couplings extracted from Table~\ref{tab:Jvalues} as a function
of the in-plane lattice parameter $a_s$. For clarity, $J'_\perp$ and $J'_{xy}$ are
scaled by a factor of $4$ and $2$, respectively, highlighting the point near
$a_s \approx 3.9$~\AA, where $E_G = E_C$. Results corresponding to the volume-relaxed
tetragonal structure ($a_s = 3.73~$\AA\ and $c_o = 4.88$~\AA) are also shown in panels
(c) and (d) using open symbols. The G-type AF and C-type AF phases are
labeled G-AF and C-AF, respectively in all figures.
}
\label{fig03}
\end{figure}

The crystal structure of $\rm BiFeO_3$ features a rhombohedral perovskite crystal
structure, which is commonly referenced as pseudocubic in this case given its
optimized reference lattice parameters ($a = b = c = 3.99$~\AA\ and $\alpha = \beta = \gamma=89.34^\circ$)~\cite{Goswami}.
This structure exhibits a G-type AF phase below 640 K~\cite{Yang}. To systematically
analyze the effects of compressive or tensile strain on this model system imposed
by a substrate we consider cubic-symmetry-based structures. For each chosen in-plane
lattice parameter ($a_s = b_s$), the out-of-plane lattice constant ($c_s$) is varied
independently. The value $c_o$ (optimized $c_s$ parameter), which is the specific
$c_s$ that minimizes the total energy of the system, is then calculated. This
procedure effectively captures the tetragonal distortion induced by epitaxial
strain. For instance, in Fig.~\ref{fig03}(a), we fixed the in-plane lattice parameter
$a_s = 4.0$~\AA\ and varied $c_s$ to find the minimum energy. Comparing the energies
of the G-type AF, C-type AF, A-type AF, and FM phases, the G-type AF and C-type
AF($\pi,\pi,0$) structures are consistently found to be lower in energy, indicating
that the main competition is between these two magnetic states. The total energies
per Fe atom of these configurations show the lowest energy achieved by the system
occurs near optimal $c_o \approx 4.2$~\AA, where the G-type AF state is the ground
state. In contrast, for a smaller in-plane lattice constant ($a_s = 3.8$~\AA), as
shown in Fig.~\ref{fig03}(b), the magnetic ground state shifts to the C-type
AF($\pi,\pi,0$) phase for an optimal $c_o = 4.75$~\AA. Please note that the
equilibrium ground state for the tetragonal structure, as established by the volume
relaxation calculation, is characterized by lattice constants $a_s = 3.73$~\AA\
and $c_o = 4.88~$\AA\ and exhibits a stable C-type AF magnetic ordering.

Next, we evaluated the energy difference $\Delta E = E_G - E_C$ between G-type AF 
and C-type AF($\pi,\pi,0$)  configurations, systematically varying the in-plane
lattice constant while optimizing the out-of-plane constant. Our results show
a clear strain-dependent trend: as the in-plane lattice constant is reduced
(compressive strain) from $a_s = 4.0$~\AA, $|\Delta E|$ steadily decreases and beyond
a critical compression, it changes sign from negative to positive. This confirms
that compressive in-plane strain enhances the tendency toward C-type AF($\pi,\pi,0$)
order. We also observed that $|\Delta E|$ increases when $a_s$ is slightly increased 
toward $a_s \approx 4.05$~\AA\ (tensile strain) from $a_s \approx 4.0$~\AA.

Now, to quantitatively describe the microscopic magnetic interactions, we employ
the following Heisenberg spin Hamiltonian:
\begin{align}
H = E_0 + J \sum_{\langle i,j \rangle} \mathbf{s}_i \cdot \mathbf{s}_j + J' \!\! \sum_{\langle\langle i,j \rangle\rangle} \!\! \mathbf{s}_i \cdot \mathbf{s}_j ,
\end{align}
where $J$ and $J'$ represent the NN and NNN exchange interactions, respectively.
We assume classical spins and set the magnitude $|\mathbf{s}| = 1$, thus absorbing
the magnitude of the moment in the exchange parameters. To capture the strain-induced
anisotropy, we distinguish between nonequivalent crystallographic directions: the
in-plane NN exchange is $J_x = J_y$, the out-of-plane NN exchange is $J_z$, the
NNN exchange along the $xz/yz$ planes is $J'_\perp$, and the in-plane NNN exchange
is $J'_{xy}$. The total energies of different spin configurations calculated
using DFT are mapped onto this model as follows (energies are given per Fe atoms):
\begin{align*}
E_G    &= E_0 - 2J_x - J_z + 4J'_\perp + 2J'_{xy}, \\
E_C    &= E_0 - 2J_x + J_z - 4J'_\perp + 2J'_{xy}, \\
E_A    &= E_0 + 2J_x - J_z - 4J'_\perp + 2J'_{xy}, \\
E_F    &= E_0 + 2J_x + J_z + 4J'_\perp + 2J'_{xy}, \\
E_{A'} &= E_0        + J_z             - 2J'_{xy},
\end{align*}
to determine five parameters $E_0$, $J_x$, $J_z$, $J'_\perp$, $J'_{xy}$. Here
$E_G$, $E_C$, $E_A$, $E_{A'}$, and $E_F$ correspond to the G-type AF, C-type
AF($\pi,\pi,0$), A-type AF($0,0,\pi$), A-type AF($0,\pi,0$), and FM states,
respectively (see Fig.~\ref{fig01} for schematic illustrations of the magnetic
phases). The extracted exchange couplings, listed in Table~\ref{tab:Jvalues},
reveal clear anisotropic trends induced by epitaxial strain. Specifically, as
the in-plane lattice constant $a_s$ decreases (and $c_o$ increases), the out-of-plane
coupling $J_z$ decreases substantially. Concurrently, the in-plane NNN coupling
$J'_{xy}$ increases as $a_s$ decreases, while the out-of-plane NNN coupling
$J'_\perp$ varies only weakly with strain. These significant anisotropies are
crucial in determining which of the competing AF states is the magnetic ground
state.

\begin{table}[t]
\centering
\caption{Direction-dependent exchange couplings ($J_x = J_y$, $J_z$, $J'_\perp$, and $J'_{xy}$)
in $\rm BiFeO_3$ obtained from DFT calculations at different in-plane lattice constants $a_s$. 
The corresponding out-of-plane lattice constant $c_o$ is the optimized value for that
particular $a_s$ (e.g., $c_o = 4.2$~\AA\ when $a_s = 4.0$~\AA). Exchange couplings for
$a_s=3.73$~\AA\ and $c_o=4.88$~\AA\ (volume-relaxed tetragonal structure)
are also provided. All $J$ values are given in meV.
}
\label{tab:Jvalues}
\begin{tabular}{c @{\hspace{0.6cm}} c @{\hspace{0.6cm}} c @{\hspace{0.6cm}} c @{\hspace{0.6cm}} c @{\hspace{0.6cm}} c}
\hline\hline
$a_s$ (\AA) & $c_o$ (\AA) & $J_x = J_y$ & $J_z$ & $J'_\perp$ & $J'_{xy}$ \\
\hline
4.05 & 4.10  & 30.93 & 35.22 & 1.77 & 1.11 \\
4.00 & 4.20  & 30.70 & 28.48 & 1.80 & 1.27 \\
3.95 & 4.40  & 27.86 & 16.29 & 1.86 & 1.56 \\
3.90 & 4.65  & 25.44 & 6.67  & 1.76 & 2.00 \\
3.85 & 4.70  & 27.28 & 5.31  & 1.74 & 2.26 \\
3.80 & 4.75  & 29.15 & 4.07  & 1.73 & 2.57 \\
3.73 & 4.88 & 30.70 & 2.21  & 1.66 & 3.19 \\
3.70 & 4.95  & 31.89 & 1.72  & 1.61 & 3.65 \\
3.60 & 5.00  & 35.87 & 1.29  & 1.55 & 4.85 \\
\hline\hline
\end{tabular}
\end{table}

A more quantitative picture of the strain effects  emerges from Fig.~\ref{fig03}(d), 
in which we plot the exchange couplings extracted via the DFT mapping (from Table~\ref{tab:Jvalues}).
For clarity in analyzing the dominant interactions, the out-of-plane NNN exchange
$J'_\perp$ is scaled by a factor of four to highlight the crossover between $4 \times J'_\perp$
and $J_z$ near $a_s \approx 3.9$~\AA, which is precisely where the energies of the
G-type AF and C-type AF($\pi,\pi,0$) states nearly balance, as shown in Fig.~\ref{fig03}(b).
This reflects the direct competition between these two exchange mechanisms given the
in-plane AF configuration: A large $J_z$ promotes the G-type AF order by stabilizing AF
alignment between out-of-plane NN Fe ions, while a larger $4 \times$ $J'_\perp$ tends to
stabilize the C-type AF($\pi,\pi,0$) phase as it favors AF coupling between NNN Fe ions
along the out-of-plane direction. The factor of four arises because each Fe ion has
eight out-of-plane NNNs but only two NNs along the $z$-direction, meaning the NNN
contribution is weighted by a factor of four when comparing to the NN contribution to
the total energy difference between G-type AF and C-type AF states. When $4J'_\perp > J_z$,
the NNN contribution becomes dominant, and the C-type AF($\pi,\pi,0$) order overtakes
the G-type AF state as the preferred magnetic configuration. Thus, the crossover
explicitly reflects the competition between the weakening of the NN out-of-plane
AF exchange ($J_z$) and the strengthening of the NNN out-of-plane exchange ($J'_\perp$)
under strain. Finally, please note that the in-plane couplings satisfy $2J'_{xy} < J_x$
throughout the studied strain range, indicating the persistent dominance of the
in-plane NN AF interaction.

\begin{table}[t]
\centering
\caption{Normalized hopping amplitudes $t_x = t_y$, $t_z$, $t'_\perp$ and $t'_{xy}$
for $\rm BiFeO_3$ are derived from the exchange couplings $J_x = J_y$, $J_z$, $J'_\perp$
and $J'_{xy}$, respectively presented in Table~\ref{tab:Jvalues}. These hopping
amplitudes are calculated using the relation $t/t_x \approx \sqrt{J/J_x}$, based
on the approximation $J \propto t^2/U$, where $U$ represents the on-site Hubbard
interaction ($t_x$ is set to $1$).}
\label{tab:t_values}
\begin{tabular}{c @{\hspace{0.7cm}} c @{\hspace{0.7cm}} c @{\hspace{0.7cm}} c @{\hspace{0.7cm}} c @{\hspace{0.7cm}} c}
\hline\hline
$a_s$ (\AA) & $c_o$ (\AA) & $t_x = t_y$ & $t_z$ & $t'_\perp$ & $t'_{xy}$ \\
\hline
4.05 & 4.10 & 1.00 & 1.067 & 0.239 & 0.189 \\
4.00 & 4.20 & 1.00 & 0.963 & 0.242 & 0.203 \\
3.95 & 4.40 & 1.00 & 0.765 & 0.258 & 0.237 \\
3.90 & 4.65 & 1.00 & 0.512 & 0.263 & 0.280 \\
3.85 & 4.70 & 1.00 & 0.441 & 0.253 & 0.288 \\
3.80 & 4.75 & 1.00 & 0.374 & 0.244 & 0.297 \\
3.73 & 4.88 & 1.00 & 0.268 & 0.233 & 0.322 \\
3.70 & 4.95 & 1.00 & 0.232 & 0.225 & 0.338 \\
3.60 & 5.00 & 1.00 & 0.190 & 0.208 & 0.368 \\
\hline\hline
\end{tabular}
\end{table}

The next step in parameterizing our Hubbard model, discussed in the next section,
is to translate the DFT-derived exchange couplings into hopping amplitudes using the
superexchange relationship $J \approx t^2/U$ where $U$ is the on-site repulsive
Hubbard interaction strength. The hopping parameter $t_x$ is calculated to be $0.37$~eV
given an exchange coupling $J_x$ of $30$~meV and an on-site repulsive Hubbard interaction
strength ($U$) of $4.6$~eV. Notably, the ratio of $U$ to $t_x$ is approximately $12.43$, 
which is a trademark of a strongly correlated electron system. With in-plane hopping $t_x$ 
set to unity as the reference scale, the normalized parameters are listed in Table~\ref{tab:t_values}.
Please note, for $a_s = 4.0$~\AA, $t_z$ nearly equals $t_x$ ($t_z/t_x \approx 0.963$),
indicating an almost isotropic three-dimensional hopping regime. As expected, the
out-of-plane hopping $t_z$ decreases considerably as the in-plane lattice constant
is reduced (i.e., under compressive strain), transitioning the system from a nearly
isotropic 3D regime to a highly anisotropic regime and selectively tuning the ratio
of out-of-plane to in-plane hopping amplitudes. Elaborately, for the smaller in-plane
lattice constant, for example $a_s = 3.6$~\AA, $t_z$ drops to approximately $0.19$,
showing that the out-of-plane hopping amplitude is much weaker than the in-plane
hopping amplitudes. Furthermore, the NNN hoppings, $t'_\perp$ and $t'_{xy}$, change
only moderately with strain. Notably, the value of $2 \times t'_\perp$ roughly matches
$t_z$ for $a_s = 3.9$~\AA, which is precisely the lattice constant where the G-type
and C-type AF phases compete as the ground state. This direct correspondence is a clear
result of our earlier analysis, as expected, where the related exchange couplings
$4 \times J'_\perp$ and  $J_z$ were found to be comparable at the same parameter
value. Overall, these results demonstrate that epitaxial strain selectively tunes
the ratio $t_z/t_x$. Thus, our first-principles results provide direct physical
intuition for our model analysis: epitaxial strain does not simply change the overall
bandwidth, but selectively modifies the ratio of out-of-plane to in-plane hopping
amplitudes. This induced anisotropy, which drives the magnetic transition, provides
the key motivation for our systematic study of the anisotropic Hubbard model at
large on-site repulsive Hubbard interaction strength.

\section{Model Hamiltonian and Computational Method}\label{sec_mm}

In order to investigate the strain-driven magnetotransport properties,
we consider the following form of the one-band half-filled Hubbard Hamiltonian on
a simple cubic lattice with periodic boundary conditions~\cite{Staudt, Mukherjee, Laubach, Fratino}:
\begin{align} \label{hamiltonian}
H =& -t \!\! \sum_{\left\langle i,j \right\rangle , \sigma} \!\! c_{i,\sigma}^\dagger c_{j,\sigma} -t' \!\!\!\! \sum_{\left\langle \left\langle i,j \right\rangle \right\rangle, \sigma} \!\!\!\! c_{i,\sigma}^\dagger c_{j,\sigma} + U \sum_i n_{i, \uparrow} n_{i, \downarrow} \nonumber \\
&- \mu \sum_i n_i = H_0 +H_I.,
\end{align}
Here, $t$ and $t'$ represent the NN and NNN hopping amplitudes, respectively.
$c_{i,\sigma}^\dagger$ ($c_{i,\sigma}$) creates (annihilates) an electron at site
$i$ with spin $\sigma$ and $n_i$ ($= \sum_\sigma c_{i,\sigma}^\dag c_{i,\sigma}$) is
the total number operator at site $i$. Furthermore, $U$ ($> 0$) is the on-site
repulsive Hubbard interaction strength, and $\mu$ is the chemical potential
controlling the overall carrier density of the system. Finally, the Hamiltonian
is formally split into a non-interacting quadratic part $H_0$ and an
interaction quartic term $H_I$.

To make the model tractable, we decouple the $H_I$ part using the Hubbard-Stratonovich (HS) transformation, 
which introduces auxiliary fields $\mathbf{m}_i$ and $\phi_i$ at each site. 
The vector field $\mathbf{m}_i$ and the scalar field $\phi_i$ are 
associated with spin and charge fluctuations, respectively. 
Next, we assume that the HS auxiliary fields are static in imaginary time and treat them as classical variables. 
At the saddle-point approximation, we relate the scalar auxiliary field $\phi_i$ 
to the local charge density $n_i$ by imposing $i\phi_i = \frac{U}{2}\langle n_i\rangle$, 
effectively making $\phi_i$ proportional to the average charge density ($\langle n_i\rangle$).
However, the auxiliary fields remain spatially non-uniform and experience thermal fluctuations, 
which are crucial for capturing finite-temperature magnetic and electronic phases. 
It is worth emphasizing that the HS transformation operates locally at each lattice site, 
enabling the decoupling of the on-site interaction term.
Using these approximations (for details, see Appendix~\ref{derivation_Heff}), 
we obtain the following effective spin-fermion Hamiltonian~\cite{Mukherjee, Jana, Chakraborty, Bidika, Halder, Mandal, Mandal2}:
\begin{align}\label{h_eff}
H_\text{eff} = & -t \!\! \sum_{\left\langle i,j \right\rangle , \sigma} \!\! c_{i,\sigma}^\dagger c_{j,\sigma} -t' \!\!\!\! \sum_{\left\langle \left\langle i,j \right\rangle \right\rangle, \sigma} \!\!\!\! c_{i,\sigma}^\dagger c_{j,\sigma} \nonumber \\ 
 & + \frac{U}{2} \sum_i \left( \left\langle n_i \right\rangle n_i - \mathbf{m}_i \cdot {\boldsymbol{\sigma}}_i \right) \nonumber \\ 
 &+ \frac{U}{4} \sum_i \left( \mathbf{m}_i^2 - \left\langle n_i \right\rangle^2 \right) - \mu \sum_i n_i,
\end{align}
where ${\boldsymbol{\sigma}}_i$ is the vector of Pauli matrices.

To solve the effective model, we employ the well studied s-MC method. 
For any given configuration of auxiliary fields $\{ \mathbf{m}_i \}$ and 
average charge densities $\{ \langle n_i \rangle \}$, the effective Hamiltonian
is initially diagonalized. The system is then simulated at a fixed temperature,
where each lattice site is sequentially updated using the Metropolis algorithm
to anneal the auxiliary spin fields $\{ \mathbf{m}_i \}$. The system undergoes
2000 Monte Carlo sweeps per temperature step: the first 1000 for thermal
equilibration and the remaining 1000 for measuring observables. To ensure the
system maintains the half-filled condition ($n = 1$), the charge densities
$\{ \langle n_i \rangle \}$ and the chemical potential $\mu$ are updated
self-consistently every 10 MC sweeps. Physical observables are computed from
100 statistically independent equilibrium configurations, with measurements
taken from every 10th configuration to reduce autocorrelation. We 
further average over 10 independent Monte Carlo calculations starting from 
different initial auxiliary-field configurations. To ensure
thorough equilibration, the temperature is decreased gradually from high
temperatures. Furthermore, we implement the traveling-cluster approximation
(TCA)~\cite{Kumar, Chakraborty, Bidika, Halder, Mandal, Mandal2} using a $4^3$
update cluster that allows us to perform simulations on large system sizes,
specifically $L^3 = 10^3$, which significantly helps to mitigate finite-size
effects. 

We incorporate direction-dependent hopping terms in the Hubbard model to account
for the anisotropy arising from the strain in the materials as discussed in previous
section. We subgroup the
three NN hopping parameters, $t_x$, $t_y$, and $t_z$, corresponding to hopping
along the $x$, $y$, and $z$ directions, respectively, into two distinct
groups: $t_x$ = $t_y$ and $t_z$. Additionally, we include NNN hopping parameters,
$t'_{xy}$, $t'_{yz}$, and $t'_{xz}$, corresponding to hopping in the $xy$, $yz$,
and $xz$ planes, and define $t'_\perp$ = $t'_{yz}$ = $t'_{xz}$. To maintain
consistency across our s-MC simulations, we set $t_x = t_y = 1$ and express all
other hopping parameters relative to $t_x$. For analyzing the effects of compressive
and tensile epitaxial strain, we tune the values of $t_z$, $t'_{xy}$, and $t'_\perp$.
We vary $0 \leq t_z \leq 2$, where $t_z < 1$ ($t_z > 1$) represents compressive
(tensile) strain in the $xy$-plane. For the NNN hopping terms, we consider
$0 \leq t'_{xy} \leq 1$ and $0 \leq t'_\perp \leq 2$. The basis for selecting
this range of hopping parameters was discussed previously using DFT calculations.
Furthermore, the on-site repulsive Hubbard interaction strength ($U$), the bare
bandwidth ($W$), and the temperature ($T$) are measured relative to the hopping
amplitude ($t_x$). This ensures that the energy scales of the interaction and
thermal effects are consistently defined relative to the hopping processes.

To gain deeper insight into the magnetotransport properties of the system, we
evaluate a range of physical observables. The magnetic behavior is characterized
by calculating the local magnetic moments $M$ and the spin structure factor $S(\mathbf{q})$,
which captures quantum spin correlations. Representative magnetic spin configurations
on a simple cubic lattice, characterized by distinct ordering vectors $\mathbf{q}$,
are shown in Fig.~\ref{fig01}. We also examine the temperature dependence of
the specific heat $C_v$, looking for low-temperature peaks that signals the magnetic
ordering phenomena. Transport characteristics are analyzed through the resistivity
along the $z$ direction $\rho_z$, defined as the inverse of the $dc$ limit of the
optical conductivity, along with the electronic density of states (DOS). A detailed
description of these observables and their computational implementation is provided in
Appendix~\ref{obs}.

\begin{figure}[t]
\centering
\includegraphics[width=0.48\textwidth]{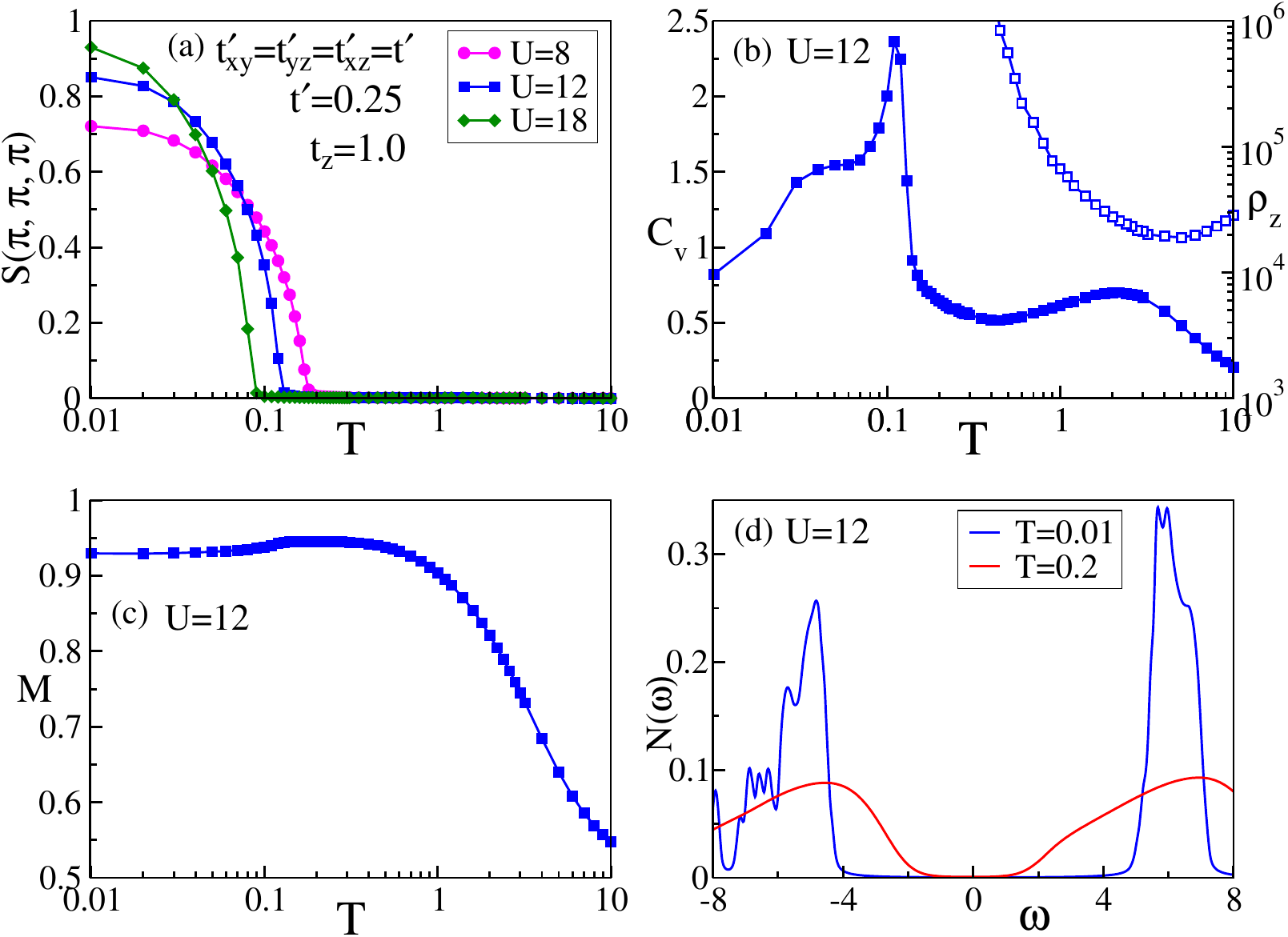}
\caption{(a) Temperature dependent of $S(\pi,\pi,\pi)$ for isotropic NN
($t_x = t_y = t_z = 1$) and NNN hoppings ($t'_{xy} = t'_{yz} = t'_{xz} = 0.25$).
The system undergoes a G-type AF transition with $T_N \approx 0.18$, $0.13$, and
$0.09$ for $U = 8, 12$, and $18$, respectively
(see main text for details). (b) Temperature dependence of the specific heat $C_v$
(left axis, solid symbols) and resistivity along the $z$ direction $\rho_z$
(right axis, open symbols) for $U = 12$. The low-$T$ peak in $C_v$ marks the onset
of long-range magnetic order, while the broad high-$T$ peak signals local-moment
formation and coincides with the metal--insulator transition temperature
$T_{\text{MIT}}$. (c) Temperature evolution of the local magnetic moment $M$ for
$U = 12$; the magnitude of $M$ grows continuously upon cooling and saturates,
reflecting the formation of localized magnetic moments at low temperatures.
(d) DOS at $T = 0.01$ and $T = 0.2$, both showing a clear gap at the Fermi level
($\omega = 0$). While the gap is wider at $T = 0.01$, it remains distinctly visible
even at $T = 0.2 > T_N$, confirming the robustness of the insulating state well
above the magnetic ordering temperature. In all calculations, the DOS is shifted
such that $\omega = 0$ corresponds to the Fermi level (chemical potential).
}
\label{fig04}
\end{figure}

We set the on-site repulsive Hubbard interaction strength $U$ to be equal to
$W$ ($\approx 12$), the bare bandwidth. This choice of $U$ value is consistent with
the DFT calculation in the previous section, where the ratio of $U$ and $t_x$ was
found to be 12.43. To further justify this parameter choice, we calculate the N\'eel
temperature $T_N$ from the temperature dependence of the magnetic structure factor
$S(\pi,\pi,\pi)$, shown in Fig.~\ref{fig04}(a) and compare with experimental results.
For isotropic NN hopping ($t_x = t_y = t_z = 1$) and isotropic NNN hopping
($t'_{xy} = t'_{yz} = t'_{xz} = 0.25$) the system undergoes a G-type AF transition
at $T_N \approx 0.13$ for $U = 12$. These parameters are consistent with the exchange
constants extracted for $a_s = 4.0$ and $c_o = 4.2$ in the DFT calculations (see
Table~\ref{tab:t_values}). Using the DFT-derived hopping amplitude $t_x \approx 0.37$~eV,
this corresponds to a N\'eel temperature of approximately 560~K, which reasonably matches
the experimental value reported for bulk $\mathrm{BiFeO_3}$ ($T_N \approx 640$~K). For
comparison, $U = 8$ and $U = 18$ [also plotted in Fig.~\ref{fig04}(a)] provide estimates
of $T_N \approx 770$~K and $T_N \approx 390$~K, respectively, both significantly
deviating from the experimentally measured value. Consequently, we set $U = 12$ for
the remainder of this work.

For completeness, we plot various other physical observables using $U = 12$ with
$t_x = t_y = t_z = 1$ and $t'_{xy} = t'_{yz} = t'_{xz} = 0.25$. The specific heat $C_v$,
shown in Fig.~\ref{fig04}(b), exhibits two distinct features: a sharp low-temperature
peak that coincides with the magnetic transition at $T_N$, and a broad high-temperature
peak that signals the formation of local magnetic moments~\cite{Paiva}. The separation of these two
peaks highlights that moment formation and long-range magnetic ordering occur on different
temperature scales, a characteristic feature of strongly correlated electronic systems.
The high-temperature peak coincides with the metal-insulator transition temperature
($T_{\mathrm{MIT}}$) extracted from the resistivity $\rho_z$ [also plotted in Fig.~\ref{fig04}(b)],
indicating that the establishment of local moments is closely tied to the opening of
an electronic gap. The temperature dependence of the local magnetic moment, $M$, shown
in Fig.~\ref{fig04}(c), further confirms that moment formation happens well above the
onset of long-range AF order and eventually saturates at low temperatures. 
The DOS obtained at $T = 0.01$ in Fig.~\ref{fig04}(d) reveals a clear gap at the Fermi
level ($\omega = 0$), confirming the insulating nature of the ordered ground state. 
The gap remains clearly visible even at $T = 0.2$, which is larger than $T_N$. The 
emergence of a gap at low temperature, in combination with the thermodynamic and transport
signatures at intermediate and large temperatures, establishes that the Mott insulating
character of the system persists across the AF transition. This confirms $T_{\mathrm{MIT}} \gg T_N$
and ensures a wide PM insulating (PM-I) regime at intermediate temperatures.

\section{Effect of anisotropic hopping: Modelling the compressive strain} \label{sec_compressive_strain}

\begin{figure}[t]
\centering
\includegraphics[width=0.48\textwidth]{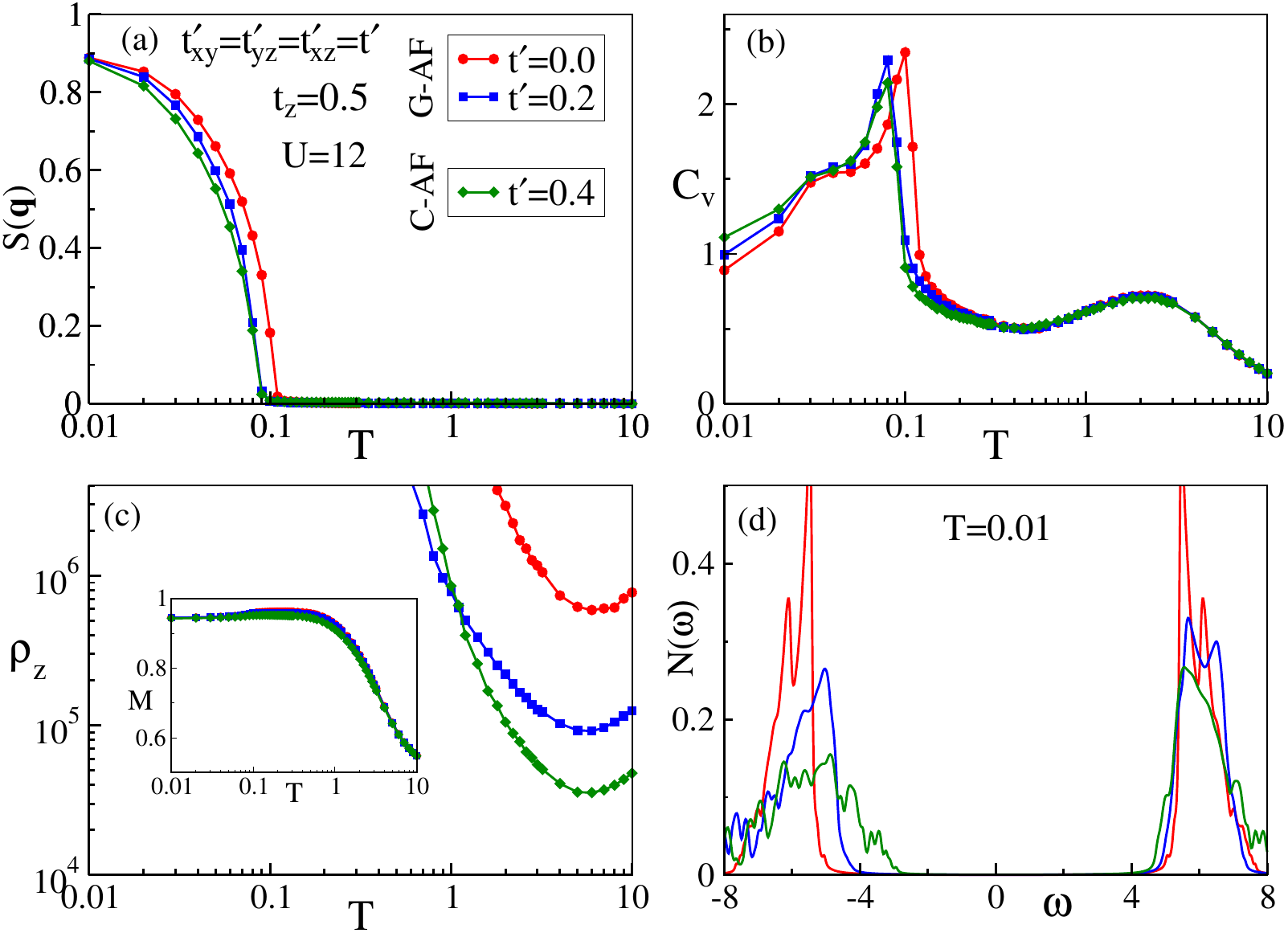}
\caption{Magnetic and transport properties for $t_z = 0.5$, with varying isotropic
NNN hopping $t' = 0.0, 0.2,$ and $0.4$ ($t' = t'_{xy} = t'_{yz} = t'_{xz}$):
(a) Magnetic structure factor $S(\mathbf{q})$ indicating G-type AF($\pi,\pi,\pi$)
ordering for $t' = 0$ and $0.2$, and C-type AF($\pi,\pi,0$) order at 
$t' = 0.4$. (b) Specific heat $C_v$ displaying low-$T$ peaks that align with the
magnetic transitions in (a), and high-$T$ peaks related to moment formation near
$T_{\text{MIT}}$. (c) Resistivity $\rho_z$ showing insulating behavior for all $t'$.
Notably, $T_{\text{MIT}} \gg T_N$, leading to a PM-I regime between the paramagnetic
metal (PM-M) and magnetically ordered phases. Inset: Temperature dependence of the
local magnetic moment $M$, which saturates at low $T$, reflecting moment formation
well above $T_N$. (d) DOS at $T=0.01$ confirming insulating ground states with a
finite gap at $\omega = 0$. Legends are consistent across all panels.
}
\label{fig05}
\end{figure}

The comprehensive DFT calculations, summarized by the derived hopping amplitudes in
Table~\ref{tab:t_values}, reveal the critical role of hopping anisotropy---the key mechanism
through which epitaxial strain alters magnetic ordering in perovskite oxides such as
$\mathrm{BiFeO_3}$. Strain primarily modifies the relative strengths of the in-plane
($t_x$, $t_y$) and out-of-plane ($t_z$) hoppings, thereby directly influencing the
competition between different AF correlations. To uncover the microscopic evolution
of magnetotransport properties under this anisotropic hopping, we examine finite-temperature
observables for $U = 12$ while fixing the out-of-plane NN hopping to $t_z = 0.5$ (mimicking
the compressive strain) and varying the isotropic NNN hopping $t'$. The magnetic
structure factors shown in Fig.~\ref{fig05}(a) clearly distinguish the competing
magnetic states. For small $t'$ ($= 0.0$ and $0.2$), $S(\pi,\pi,\pi)$ exhibits a
well-defined transition into a G-type AF state, with the corresponding
N\'eel temperature $T_N$ decreasing slightly as $t'$ increases. This decrease reflects
enhanced magnetic frustration introduced by NNN hopping, which weakens the
three-dimensional AF correlations. For larger $t'$ ($= 0.4$), the magnetic
correlations shift to $S(\pi,\pi,0)$, signaling a crossover from G-type AF to
C-type AF($\pi,\pi,0$) order.

The specific heat $C_v$ curves, shown in Fig.~\ref{fig05}(b), display sharp
low-temperature peaks, which track the magnetic transitions identified in
Fig.~\ref{fig05}(a). These peaks, related to spin fluctuations, thermodynamically
verify the ordering temperatures. A broad high-temperature peak, originating
from charge fluctuations, persists across all $t'$, indicating the robust
formation of local magnetic moments~\cite{Paiva}. The temperature dependence of the local
magnetic moment, $M$, shown in the inset of Fig.~\ref{fig05}(c), further confirms
that moments are well formed at low temperatures for all three $t'$ values.
The main panel of Fig.~\ref{fig05}(c)  shows that the resistivity along the
$z$ direction, $\rho_z$, remains insulating at low temperatures for all $t'$
values, although $T_{\mathrm{MIT}}$ decreases slightly with increasing $t'$.
This observed insulating behavior is consistently supported by the low-temperature
DOS in Fig.~\ref{fig05}(d), which reveals a gap around the Fermi level ($\omega = 0$).
Furthermore, we note that $T_{\mathrm{MIT}}$ matches
well with the temperature at which the high-temperature peak in $C_v$ appears
[as discussed for the isotropic hopping case, in Fig.~\ref{fig04}(b)]. Importantly,
$T_{\mathrm{MIT}}$ always exceeds $T_N$, which ensures the existence of 
a substantial PM-I regime between the PM-M and AF insulating states.

\begin{figure}[t]
\centering
\includegraphics[width=0.48\textwidth]{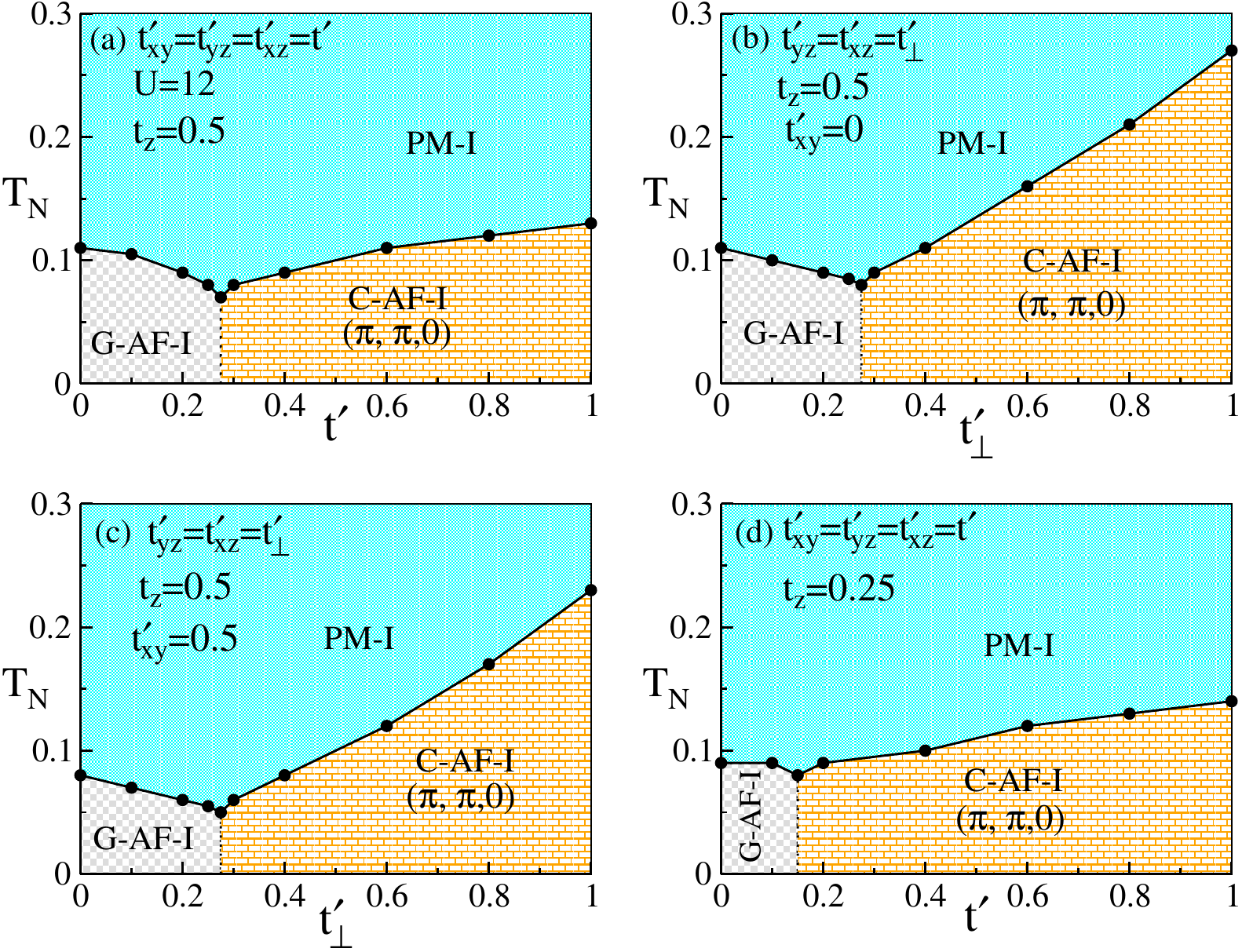}
\caption{(a) The $t'$-$T_N$ phase diagram for isotropic NNN hopping
($t'_{xy} = t'_{yz} = t'_{xz} = t'$) at $t_z = 0.5$. Small $t'$ favors a G-type AF
insulating phase, while moderate-to-large $t'$ stabilizes a C-type 
AF($\pi,\pi,0$) insulating phase. As $t'$ increases, $T_N$ decreases within
the G-type AF regime but increases for the C-type AF phase. The $t'_{\perp}$-$T_N$
phase diagrams ($t'_{yz} = t'_{xz} = t'_{\perp}$) with $t_z = 0.5$ for two in-plane
NNN hoppings: (b) $t'_{xy} = 0$ and (c) $t'_{xy} = 0.5$. In both cases, increasing
$t'_{\perp}$ drives a transition from G-type AF at small $t'_{\perp}$ to C-type AF
at larger values. The transition point is essentially identical to that in (a),
suggesting the in-plane NNN hopping does not shift the crossover boundary,
though $T_N$ is suppressed for $t'_{xy} = 0.5$. (d) The $t'$-$T_N$ phase diagram
for isotropic NNN hopping at a smaller out-of-plane hopping $t_z = 0.25$. Reducing
$t_z$ shifts the G-type AF to C-type AF crossover to lower $t'$ values. In all cases,
$T_{\text{MIT}} \gg T_N$, giving rise to a PM-I regime above the $T_N$. The G-type
and C-type AF phases are labeled as G-AF-I and C-AF-I, respectively; notably,
all magnetic phases reported in this work are insulating.
}
\label{fig06}
\end{figure}

We now map out the $t'$-$T_N$ phase diagram for isotropic NNN hopping ($t'_{xy}=t'_{yz}=t'_{xz}=t'$)
in Fig.~\ref{fig06}(a). As the temperature is lowered, the system evolves from the high-temperature
PM-I regime, transitioning into a G-type AF insulator for small $t'$ but stabilizing a C-type AF($\pi,\pi,0$) 
insulating state for moderate to large $t'$. Increasing $t'$ monotonically suppresses
the G-type AF ordering temperature ($T_N$) due to the enhanced magnetic frustration introduced
by the NNN hopping. This G-type AF order persists until a critical value of $t' \approx 0.27$,
beyond which the system stabilizes the C-type AF($\pi,\pi,0$) phase, and $T_N$ subsequently
increases with further increments in $t'$. Crucially, the anisotropy in the NN hopping
($t_x = t_y \neq t_z$) explicitly lifts the degeneracy of the C-type AF orders [i.e., between
the $(\pi,\pi,0)$, $(0,\pi,\pi)$, and $(\pi,0,\pi)$ variants] that arises from cubic symmetry.
Since $t_z < t_x$, the out-of-plane AF order is weaker than the in-plane AF order; consequently,
only the C-type AF($\pi,\pi,0$) configuration is stabilized in the phase diagram for large $t'$,
while the symmetry-related $(0,\pi,\pi)$ and $(\pi,0,\pi)$ variants are suppressed.
Throughout this evolution, $T_{\mathrm{MIT}}$ remains higher than $T_N$, which maintains
a broad PM-I intermediate regime, consistent with the magnetic and transport behavior
reported in Fig.~\ref{fig05}.

At this point it is important to mention that while the isotropic $t'$ analysis identifies
the dominant role of NNN hopping in controlling the magnetic phase competition, realistic
epitaxial strain environments are expected to modify the in-plane and out-of-plane NNN
hopping amplitudes differently. This raises a critical question: which NNN hopping component---the
out-of-plane component ($t'_{\perp}=t'_{yz}=t'_{xz}$) or the in-plane component ($t'_{xy}$)---plays
the major role in driving the magnetic phase transition? To address this, we analyze the
$t'_{\perp}$-$T_N$ phase diagrams for anisotropic NNN hopping configurations shown in Figs.~\ref{fig06}(b)
and \ref{fig06}(c).  In these plots, the out-of-plane component $t'_{\perp}$ is varied at a
fixed $t_z = 0.5$ (compressive strain) for two representative in-plane values: $t'_{xy}=0$
[Fig.~\ref{fig06}(b)] and $t'_{xy}=0.5$ [Fig.~\ref{fig06}(c)]. In both cases, increasing $t'_{\perp}$
drives a transition from a G-type AF state to a C-type AF($\pi,\pi,0$) phase. Remarkably, the
transition point for the G-type AF to C-type AF($\pi,\pi,0$) crossover is essentially identical
to that found in the isotropic NNN hopping case [Fig.~\ref{fig06}(a)]. This strongly suggests
that the crossover boundary is governed primarily by the out-of-plane NNN hopping $t'_{\perp}$
rather than the in-plane component $t'_{xy}$, a result that is physically viable in real
systems given that the required $t'_\perp$ is around $t_z$/2 to drive this transition.

However, $t'_{xy}$ does influence the ordering
temperature: when the in-plane competition is absent ($t'_{xy} = 0$), the $T_N$ corresponding
to the C-type AF($\pi,\pi,0$) phase grows rapidly with increasing $t'_\perp$. Conversely, when
a finite in-plane NNN hopping is included ($t'_{xy} = 0.5$), the overall N\'eel temperature
decreases across all $t'_\perp$. This confirms that the in-plane NNN hopping $t'_{xy}$ mainly
modulates the ordering temperature under compressive strain ($t_z = 0.5$).

Next, we investigate the isotropic NNN hopping case by setting the out-of-plane NN hopping to
$t_z = 0.25$ to focus on the effects of stronger compressive strain. This allows us to
quantify how strain shifts the balance between competing AF orders. The resulting $t'$-$T_N$
phase diagram is displayed in Fig.~\ref{fig06}(d).  Under these conditions, the transition
from G-type AF to C-type AF($\pi,\pi,0$) occurs at a smaller critical $t'$ ($\approx 0.15$),
confirming that compressive strain enhances the tendency toward C-type AF($\pi,\pi,0$) order
at much smaller $t'$. Similar to the behavior observed at $t_z = 0.5$, in this more extreme
compressive regime ($t_z = 0.25$), the primary role of the in-plane NNN hopping, $t'_{xy}$,
is to modulate the magnetic ordering temperature. However, crucially, $t'_{xy}$ does not
significantly control the magnetic phase transition point at low temperatures, which is
predominantly controlled by the out-of-plane NNN hopping, $t'_{\perp}$.

\begin{figure}[t]
\centering
\includegraphics[width=0.48\textwidth]{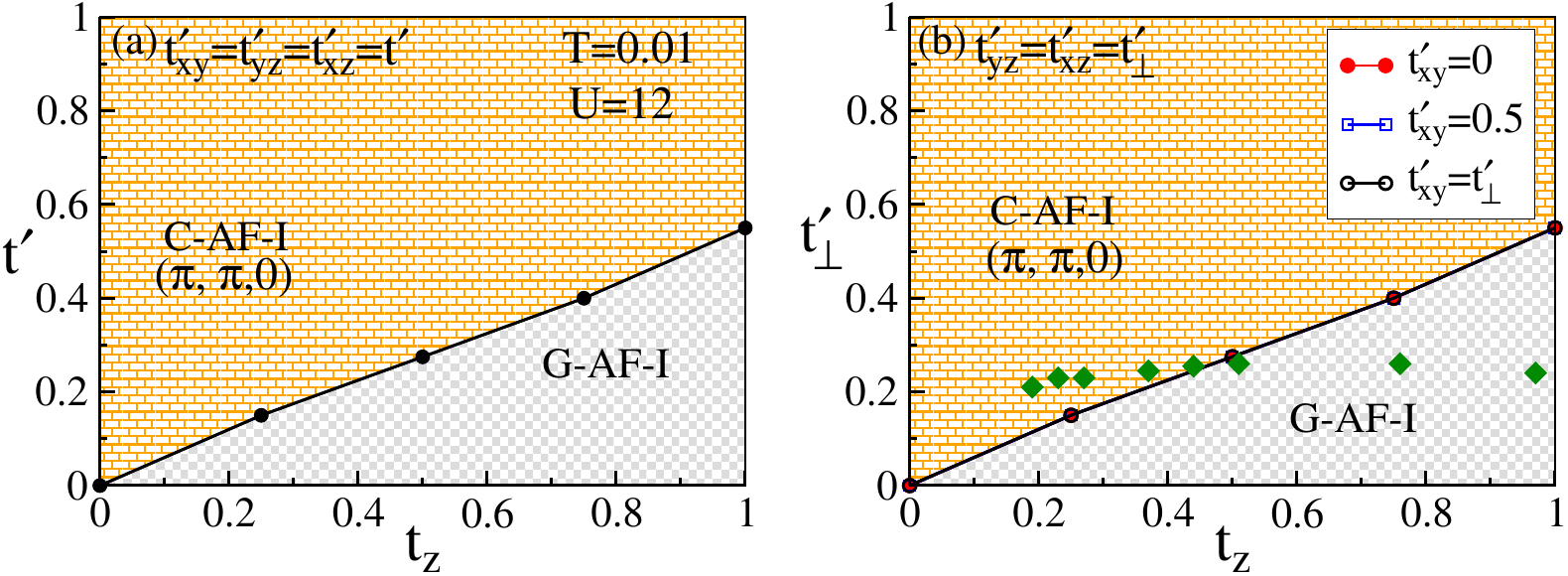}
\caption{(a) Low-temperature ($T = 0.01$) $t_z$-$t'$ phase diagram for isotropic
NNN hoppings. Increasing $t'$ drives a transition from the G-type AF insulating
state to a C-type AF insulating state. The critical $t'$ value increases linearly
as $t_z$ increases.
(b) The $t_z$-$t'_{\perp}$ phase diagrams ($t'_{yz} = t'_{xz} = t'_{\perp}$) shown
for $t'_{xy} = 0$ and $t'_{xy} = 0.5$; the isotropic NNN hopping case
($t'_{xy} = t'_{\perp}$) is replotted for comparison. In all cases, increasing
$t'_{\perp}$ destabilizes the G-type AF phase in favor of C-type AF order, while the
phase boundary remains independent of $t'_{xy}$. Green diamonds mark
the parameter sets obtained from DFT (see Table~\ref{tab:t_values}).}
\label{fig07}
\end{figure}

The finite-temperature analysis above demonstrates that the competition between G-type
AF and C-type AF($\pi,\pi,0$) orders is strongly controlled by the strengths of both
out-of-plane NN and NNN hopping amplitudes. To present the complete picture of how the
ground state phases behave with $t_z$ and $t'$, we show the $t_z$-$t'$ phase diagram
at a low temperature ($T = 0.01$) in Fig.~\ref{fig07}(a). Note that, in this section, our analysis
focuses only on the $t_z < t_x$ regime, which corresponds to compressive strain. As
the isotropic NNN hopping $t'$ increases, the system evolves from a G-type AF insulator
to a C-type AF($\pi,\pi,0$) insulator in the whole range. The critical $t'$ separating
these phases grows monotonically with increasing $t_z$, as a larger out-of-plane NN
hopping $t_z$ stabilizes AF alignment along the $z$-axis and delays the ferromagnetic
stacking required for the C-type AF($\pi,\pi,0$) state.

Next, focusing on the relative roles of in-plane and out-of-plane NNN hopping, we show
the $t_z$-$t'_\perp$ phase diagram at $T = 0.01$ with varying $t'_{xy}$ in
Fig.~\ref{fig07}(b). In all cases shown ($t'_{xy} = 0$, $t'_{xy} = 0.5$, and $t'_{xy} = t'_\perp$),
increasing the out-of-plane NNN hopping $t'_\perp$ destabilizes the G-type AF phase and
stabilizes the C-type AF($\pi,\pi,0$) order. Notably, the phase boundary between the G-type
AF and C-type AF($\pi,\pi,0$) order remains nearly unchanged across all three $t'_{xy}$
values. This highlights that for $t_z < t_x$ (under compressive strain), the out-of-plane
NNN hopping $t'_\perp$ is the main driver of the G-type AF to C-type AF($\pi,\pi,0$)
transition, while variations in in-plane NNN hopping $t'_{xy}$ primarily influence
the ordering temperature (as seen in Fig.~\ref{fig06}).

The magnetic phase boundary in Fig.~\ref{fig07}(b), marking the transition from G-type
to C-type AF($\pi,\pi,0$), is consistently found near the hopping ratio
$t'_\perp/t_z \approx 0.5$. This result is fully consistent with the DFT-derived
crossing point, which shows the magnetic crossover is governed by $J'_\perp/J_z \approx 0.25$
(Fig.~\ref{fig03}), based on the underlying superexchange mechanism $J \sim t^2/U$. 
These $J'_\perp/J_z \approx 0.25$ or $t'_\perp/t_z \approx 0.5$ conditions signify the
point where the out-of-plane NN AF interaction is balanced by the frustrating out-of-plane NNN AF
interaction. The location of the DFT-derived hopping parameters (indicated by diamond
symbols in Fig.~\ref{fig07}(b) and listed in Table~\ref{tab:t_values}) near the G-type
AF to C-type AF($\pi,\pi,0$) boundary confirms that a small change in out-of-plane
NN hopping can switch the magnetic ground state. Overall, these results demonstrate that
the anisotropic hopping framework is minimal yet effective for capturing and
predicting the strain-induced magnetic ground state switch.

\begin{figure}[t]
\centering
\includegraphics[width=0.3\textwidth]{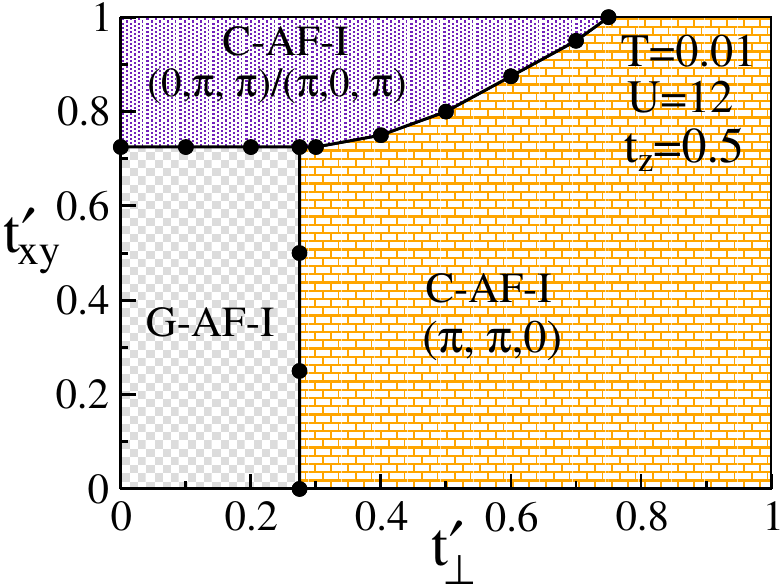}
\caption{The $t'_{\perp}$-$t'_{xy}$ phase diagram at $T = 0.01$ for $t_z = 0.5$
(representing compressive strain). At small $t'_{xy}$, increasing $t'_{\perp}$
drives a transition from the G-type AF insulating state to the C-type AF($\pi,\pi,0$)
insulating phase. Conversely, for large in-plane NNN hopping ($t'_{xy} \gtrsim 0.72$),
a twofold-degenerate C-type AF[($0,\pi,\pi$)/($\pi,0,\pi$)] phase is stabilized at small
$t'_{\perp}$. Within this large-$t'_{xy}$ regime, increasing $t'_{\perp}$ induces a
switching between the C-type AF[($0,\pi,\pi$)/($\pi,0,\pi$)] and C-type AF($\pi,\pi,0$) phases.
}
\label{fig08}
\end{figure}

Finally, to present the full magnetic phase behavior across the parameter space, we
show the $t'_\perp$-$t'_{xy}$ phase diagram in Fig.~\ref{fig08} for $t_z = 0.5$.
Here, we varied $t'_{xy}$ up to a value of 1 for completeness, though we acknowledge
that such a large magnitude relative to $t_x$ is highly improbable in realistic
material systems. Beyond a critical, high value of $t'_{xy}$ (approximately $\sim 0.72$),
the twofold-degenerate C-type AF[$(0,\pi,\pi)$/$(\pi,0,\pi)$] phase appears as the
ground state, particularly for small values of the out-of-plane NNN hopping, $t'_\perp$.
Furthermore, in this high $t'_{xy}$ regime (e.g., around $t'_{xy} \sim 0.8$), increasing
$t'_\perp$ triggers a subtle intra-C-type phase transition, causing a switching of
the ground state between the C-type variants: AF[$(0,\pi,\pi)$/$(\pi,0,\pi)$] and
AF($\pi,\pi,0$) phase.

\section{Effect of anisotropic hopping: Modelling the tensile strain} \label{sec_tensile_strain}

\begin{figure}[t]
\centering
\includegraphics[width=0.48\textwidth]{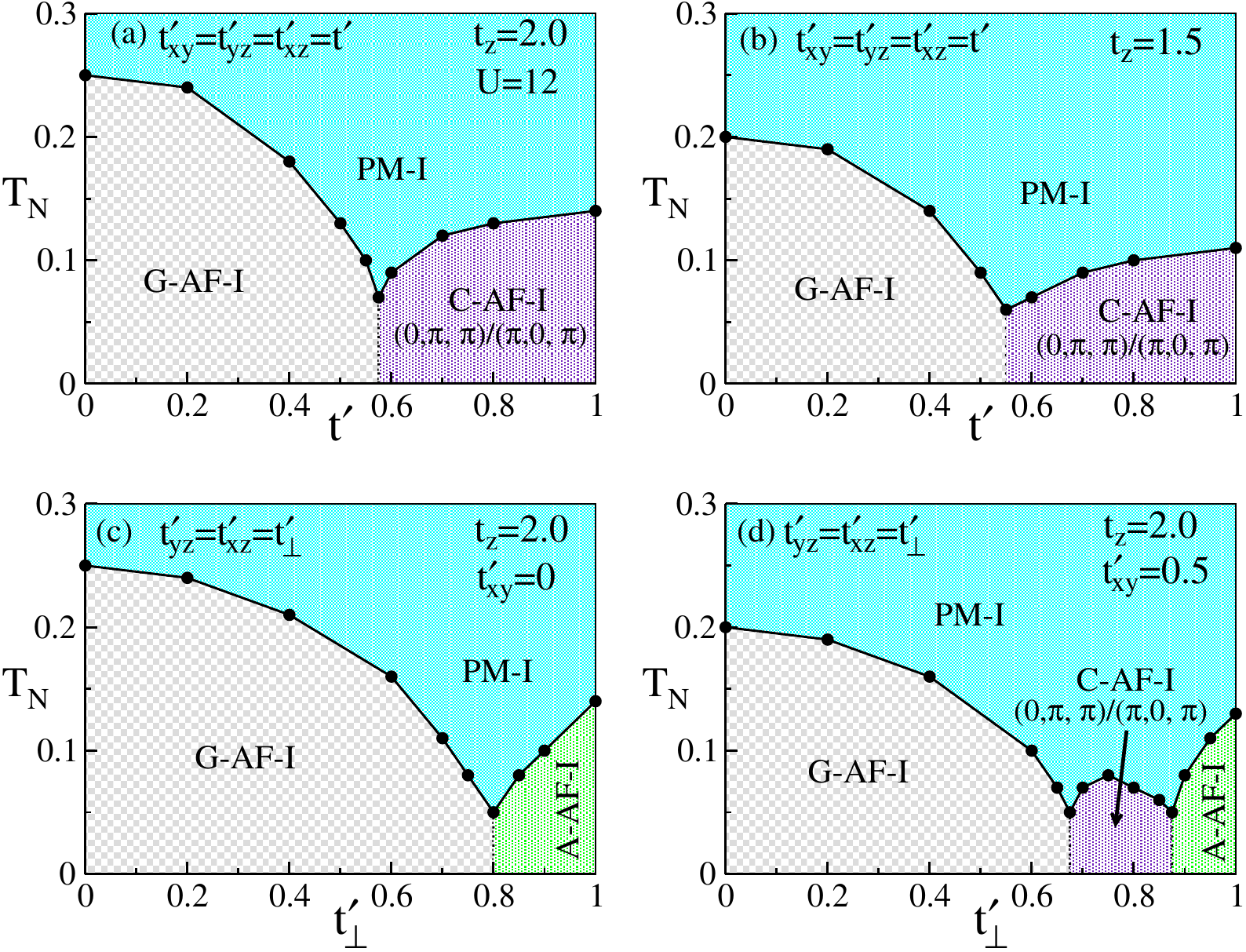}
\caption{Finite-temperature phase diagrams for isotropic and anisotropic NNN hoppings
under tensile strain ($t_z > t_x$). (a, b) The $t'$-$T_N$ phase diagrams for isotropic
NNN hoppings ($t'_{xy} = t'_{yz} = t'_{xz} = t'$) at (a) $t_z = 2.0$ and (b) $t_z = 1.5$.
For both $t_z$ values, the system orders into a G-type AF insulating state at small $t'$,
whereas larger $t'$ stabilizes a twofold-degenerate C-type AF[($0,\pi,\pi$)/($\pi,0,\pi$)]
insulating phase. Across the G-type AF region, $T_N$ decreases monotonically with increasing $t'$,
while in the C-type AF region, $T_N$ increases with $t'$. Reducing $t_z$ from 2.0 to 1.5
suppresses $T_N$ overall, though the qualitative features of the phase diagram remain
unchanged. (c, d) The $t'_{\perp}$-$T_N$ phase diagrams for anisotropic NNN hopping
($t'_{yz} = t'_{xz} = t'_{\perp}$) at $t_z = 2.0$ with (c) $t'_{xy}=0$ and
(d) $t'_{xy}=0.5$. In the absence of in-plane NNN hopping, the twofold-degenerate
C-type AF[($0,\pi,\pi$)/($\pi,0,\pi$)] phase is fully suppressed; instead, an A-type AF($0,0,\pi$) insulating
phase emerges at large $t'_{\perp}$, and the G-type AF region expands relative to (a).
With finite in-plane hopping ($t'_{xy} = 0.5$), the phase diagram becomes richer:
the G-type AF insulating phase persists up to intermediate $t'_{\perp}$, followed by a narrow
window of the twofold-degenerate C-type AF[($0,\pi,\pi$)/($\pi,0,\pi$)] phase, and finally an A-type AF($0,0,\pi$) state at
large $t'_{\perp}$. Interestingly, for $t'_{xy} = 0.5$, $T_N$ exhibits a non-monotonic
trend within the twofold-degenerate C-type AF[($0,\pi,\pi$)/($\pi,0,\pi$)] region, while it continues to
decrease monotonically for G-type AF and increase for A-type AF($0,0,\pi$) orders. In all
panels, $T_{\text{MIT}}$ remains significantly higher than $T_N$, resulting in a
finite-temperature PM-I regime. A-type AF($0,0,\pi$) insulating phase is labeled as A-AF-I
in all figures.
}
\label{fig09}
\end{figure}

Our calculations in the previous section established that compressive epitaxial strain
($t_z < t_x$) strongly enhances the competition between two types of magnetic order,
namely G-type and C-type AF order, with NNN hopping selectively stabilizing the
C-type AF($\pi,\pi,0$) state; however, the magnetic competition is expected to be
qualitatively modified under tensile strain ($t_z > t_x$), where the out-of-plane
NN hopping becomes the dominant coupling. To determine how this balance evolves, we
now examine the finite-temperature phase diagram for a representative large value,
$t_z = 2.0$, and analyze its interplay with both isotropic and anisotropic NNN hopping.
For isotropic NNN hopping, $t'_{xy} = t'_{yz} = t'_{xz} = t'$ [see Fig.~\ref{fig09}(a)],
the system shifts from a high-temperature PM-I phase into a G-type AF insulator
at small $t'$, but the ground state switches to a twofold-degenerate C-type AF phase,
corresponding to ordering vectors $(0,\pi,\pi)$ and $(\pi,0,\pi)$, at large $t'$.

The analysis of the tensile strain regime ($t_z > t_x$) reveals three primary
consequences arising from the dominant out-of-plane NN hopping. The first key
finding is the overall enhancement of the $T_N$ of the G-type AF phase, directly
attributable to the large $t_z$ which drives the AF order
to higher temperatures. The second key finding is the complete absence of the
C-type AF($\pi,\pi,0$) phase [which was present in the small-$t_z$ regime
(see Fig.~\ref{fig06})] when $t_z$ is large. This difference also stems from
the large out-of-plane NN hopping: since $t_z > t_x$, the NN AF coupling along
the $z$ direction becomes dominant over in-plane NN AF interactions. Consequently,
the system naturally favors magnetic orders that preserve out-of-plane
antiferromagnetism, leading to the stabilization of the twofold-degenerate C-type
AF[$(0,\pi,\pi)$/$(\pi,0,\pi)$] phase at large $t'$.  The third key finding is
that the G-type AF order persists up to significantly larger NNN hopping ($t'$) values
compared to the small-$t_z$ case [see Fig.~\ref{fig06}(a)]. The G-type AF phase 
maintains its stability over an extended intermediate range of $t'$ due to
two combined effects: the increasing influence of the NN hopping parameters
with larger $t_z$, and the degeneracy of the competing C-type AF phase
between the ($0,\pi,\pi$) and ($\pi,0,\pi$) ordering vectors.
Consistent with the small-$t_z$ regime, the $T_N$ decreases
monotonically with $t'$ in the G-type AF phase, while it increases
monotonically in the twofold-degenerate C-type $\text{AF}[(0,\pi,\pi)/(\pi,0,\pi)]$
phase. The qualitative scenario remains the same for $t_z = 1.5$, as shown in
Fig.~\ref{fig09}(b). 

While the analysis of the isotropic NNN hopping case effectively demonstrates that large
out-of-plane NN hopping ($t_z > t_x$) fundamentally reshapes the magnetic
competition, it does not by itself identify which NNN hopping channel is
responsible for this restructuring. In particular, large $t_z$ removes the
C-type AF($\pi,\pi,0$) phase and stabilizes the twofold-degenerate C-type
AF[($0,\pi,\pi$)/($\pi,0,\pi$)] order at large $t'$, but it remains unclear
whether this behavior is driven by the out-of-plane NNN hopping $t'_\perp$ or
by the in-plane NNN hopping $t'_{xy}$.
To disentangle these effects, we selectively remove the in-plane
NNN hopping ($t'_{xy} = 0$) and analyze the regime where only the out-of-plane
NNN hopping, $t'_\perp$, competes with the dominant out-of-plane NN hopping.
Upon switching off the in-plane NNN hopping [see Fig.~\ref{fig09}(c)], the
twofold-degenerate C-type AF[$(0,\pi,\pi)$/$(\pi,0,\pi)$] phase is completely
suppressed, and an A-type AF($0,0,\pi$) insulating state is stabilized instead
at large $t'_\perp$. This shift is physically intuitive: the large $t_z$
favors AF alignment between NNs along the $z$ direction, and a large $t'_\perp$
also promotes AF alignment between NNNs along $z$. The resulting A-type
AF($0,0,\pi$) order accommodates both tendencies simultaneously, as spins
are ferromagnetically aligned within each $xy$ plane but alternate
antiferromagnetically between adjacent planes. In this specific configuration,
both out-of-plane NN and NNN bonds are antiferromagnetically satisfied, making
the A-type AF($0,0,\pi$) phase the energetically optimal solution in this
regime. Notably, the G-type AF phase is also stabilized over a wider
$t'_\perp$ range compared to the preceding isotropic case in Fig.~\ref{fig09}(a).

For $t'_{xy} = 0.5$ [see Fig.~\ref{fig09}(d)], the resulting phase diagram
is significantly richer. The system exhibits a sequential transition: the G-type
AF phase persists up to intermediate values of $t'_\perp$, followed by a narrow
window of the twofold-degenerate C-type AF[$(0,\pi,\pi)$/$(\pi,0,\pi)$] order,
and finally transitioning to the A-type AF($0,0,\pi$) phase at large $t'_\perp$.
Within this complex landscape, $T_N$ exhibits a non-monotonic trend in the C-type
AF region, while it continues to decrease for G-type AF and increase for A-type AF,
indicating subtle competition between the different NNN couplings. The emergence
of the degenerate C-type AF[($0,\pi,\pi$)/($\pi,0,\pi$)] region is key, as it
highlights the interplay of competing exchange channels: the presence of finite
$t'_{xy}$ introduces in-plane frustration that destabilizes the G-type AF state
and promotes this intermediate C-type AF phase. Thus, the resulting phase sequence
arises from two simultaneous competitions: first, the interplay between the
out-of-plane NN ($t_z$) and NNN ($t'_\perp$) hopping, and second, the balance
between the in-plane NN ($t_x = t_y$) and NNN ($t'_{xy}$) hopping. Crucially,
these calculations demonstrate that the in-plane NNN hopping ($t'_{xy}$) plays
a decisive role in stabilizing and modulating the magnetic orders across the
phase space.

\begin{figure}[t]
\centering
\includegraphics[width=0.3\textwidth]{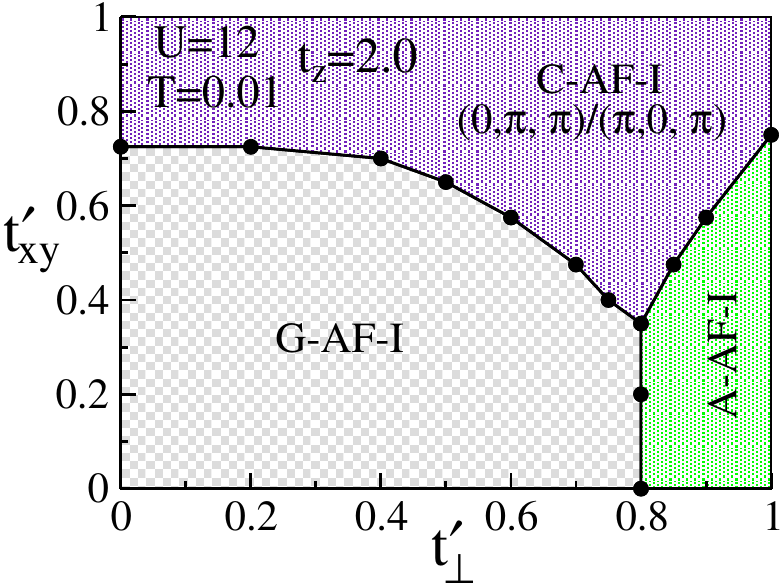}
\caption{The $t'_{\perp}$-$t'_{xy}$ phase diagram at $T = 0.01$ for $t_z = 2.0$
(representing tensile strain). For small $t'_{xy}$, the ground state consists of
G-type AF insulating state at small to moderate $t'_{\perp}$, followed by an A-type
AF($0,0,\pi$) insulating state at sufficiently large $t'_{\perp}$ ($\gtrsim 0.8$). 
As $t'_{xy}$ increases, in-plane frustration stabilizes a twofold-degenerate 
C-type AF[($0,\pi,\pi$)/($\pi,0,\pi$)] phase over a broad parameter region.
}
\label{fig10}
\end{figure}

Next, we present the $t'_\perp$-$t'_{xy}$
phase diagram in Fig.~\ref{fig10}, which clearly demonstrates that the complex
emergence and competition of magnetic phases primarily occurs when the in-plane
NNN hopping is significant ($t'_{xy} \gtrsim 0.35$). Notably, beyond a high value
of $t'_{xy}$ ($\sim 0.75$), the twofold-degenerate C-type AF[$(0,\pi,\pi)$/$(\pi,0,\pi)$]
phase dominates the ground state, irrespective of the value of the
out-of-plane NNN hopping, $t'_\perp$.  It is important to note that such high
$t'_{xy}$ values may not be physically viable in real systems. Taken together,
our calculations demonstrate that under tensile strain ($t_z = 2$), the dominance
of out-of-plane NN hopping qualitatively reshapes the magnetic competition,
leading to the emergence of distinct G-type, C-type, and A-type AF regimes,
with the aid of $t'_{xy}$ and $t'_\perp$, that were absent in the compressive
limit (e.g., the $t_z =0.5$ case).

\begin{figure}[t]
\centering
\includegraphics[width=0.48\textwidth]{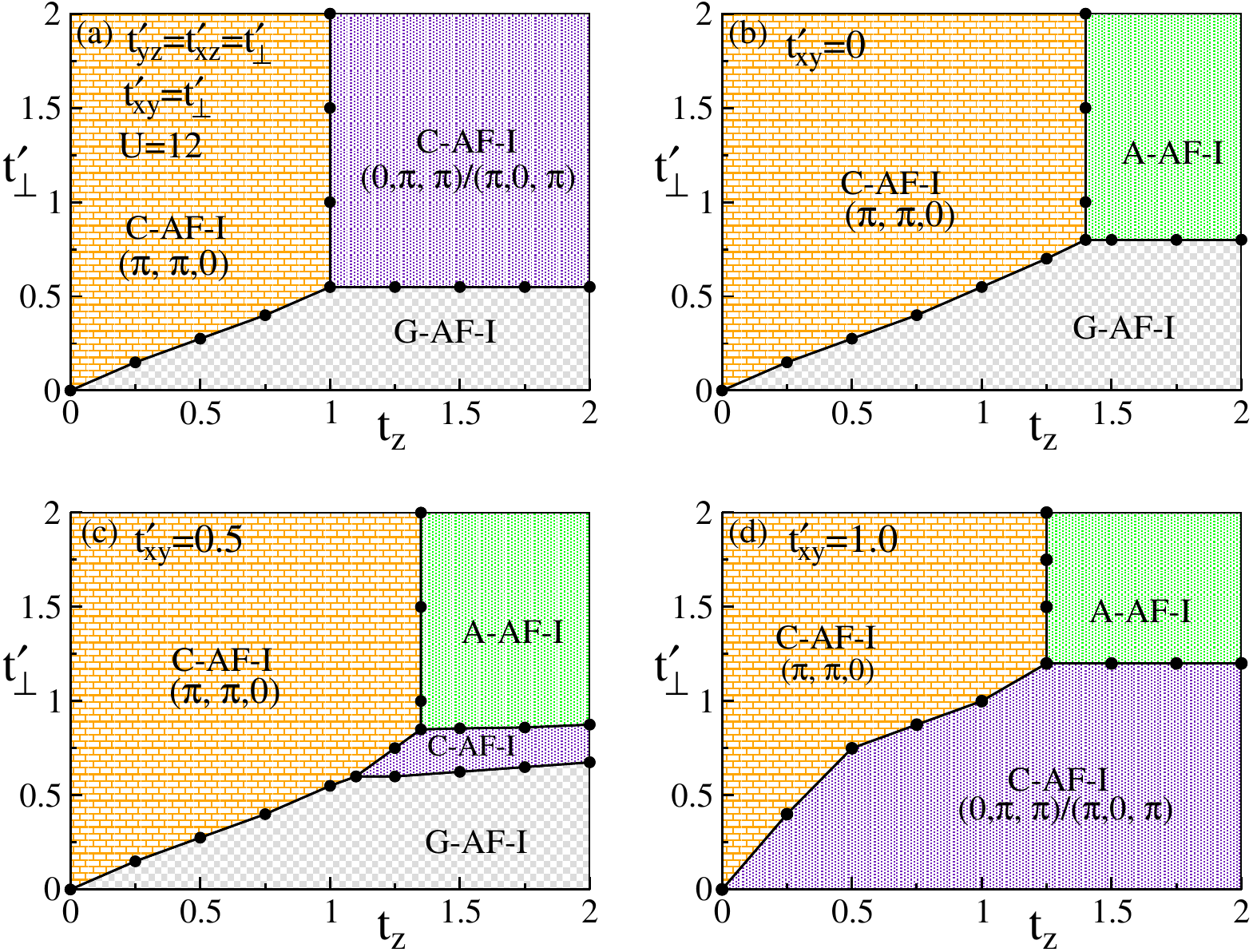}
\caption{The $t_z$-$t'_\perp$ phase diagrams at $T = 0.01$ for various $t'_{xy}$,
illustrating the evolution of magnetic ground states from the compressive ($t_z < t_x$)
to tensile ($t_z > t_x$) strain regimes. (a) Isotropic NNN hopping
($t'_{xy} = t'_{\perp}$): increasing $t'_{\perp}$ (i.e., $t'$ in this case) drives
a transition from the G-type AF phase to a C-type AF phase for all $t_z$. For
$t_z < 1$, the C-type AF($\pi,\pi,0$) state is stabilized with a critical $t'_\perp$
that increases linearly with $t_z$. For $t_z > 1$, the system favors a doubly
degenerate C-type AF[($0,\pi,\pi$)/($\pi,0,\pi$)] phase with a nearly constant
critical $t'_\perp$. At $t_z = 1$, cubic symmetry is restored, and the G-type AF insulating phase
transitions to a threefold-degenerate C-type AF[($\pi,\pi,0$)/($0,\pi,\pi$)/($\pi,0,\pi$)]
insulating state for $t'_\perp \gtrsim 0.55$. (b) Anisotropic NNN hopping with $t'_{xy} = 0$:
the critical out-of-plane hopping $t'_{\perp}$ required for the G-type AF to
C-type AF($\pi,\pi,0$) transition increases with $t_z$ up to $t_z \sim 1.4$,
beyond which an A-type AF($0,0,\pi$) phase is stabilized at larger $t'_{\perp}$.
(c) Moderate in-plane NNN hopping ($t'_{xy} = 0.5$): the low-$t_z$ behavior
is similar to that in (b); however, in the tensile strain regime, the G-type AF
phase gives way first to the doubly degenerate C-type AF[($0,\pi,\pi$)/($\pi,0,\pi$)]
phase and subsequently to the A-type AF($0,0,\pi$) phase at larger $t'_{\perp}$.
(d) Large in-plane NNN hopping ($t'_{xy} = 1$): the doubly degenerate
C-type AF[($0,\pi,\pi$)/($\pi,0,\pi$)] phase dominates the phase diagram,
suppressing the G-type AF region and significantly reducing the stability of
the A-type AF($0,0,\pi$) phase.
}
\label{fig11}
\end{figure}

Finally, for completeness, we analyze the low-temperature phase behavior over $0 \le t_z \le 2$
to present a unified understanding of how magnetic states evolve across the entire range
of hopping, from compressive ($t_z < t_x$) to tensile ($t_z > t_x$) strain in
Fig.~\ref{fig11}. While $0 \le t_z \le 1$ was previously discussed, the full range
is presented here for a detailed comparison across all magnetic phase transitions.
The resulting low-temperature $t_z$-$t'_\perp$ phase diagram, where NNN hoppings are isotropic ($t'_{xy} = t'_\perp = t'$),
is presented in Fig.~\ref{fig11}(a). For any $t_z$, increasing $t'_\perp$, a transition from the
G-type AF insulator to the C-type AF insulator is driven, as discussed earlier. Specifically,
for $t_z < 1$, this C-type phase is the AF($\pi,\pi,0$), and the critical $t'_\perp$
value increases linearly with $t_z$. Conversely, for $t_z > 1$, the phase shifts to a
doubly degenerate C-type AF[($0,\pi,\pi$)/($\pi,0,\pi$)] state, and the critical $t'_\perp$
value remains nearly constant. For $t_z = 1$, the G-type AF order transitions
to a threefold degenerate C-type AF[($0,\pi,\pi$)/($\pi,0,\pi$)/($\pi,\pi,0$)] phase
for $t'_\perp \gtrsim 0.55$, due to the restoration of cubic symmetry in the NN and
NNN hopping.

We next vary the in-plane ($t'_{xy}$) and out-of-plane ($t'_\perp$) NNN hopping
independently. For $t'_{xy} = 0$ [see Fig.~\ref{fig11}(b)], as $t_z$ increases,
the critical $t'_\perp$ value required for the G-type to C-type transition increases
linearly up to $t_z \sim 1.4$ (this was up to $t_z =1 $ for isotropic NNN hopping).
Beyond $t_z \sim 1.4$, increasing $t'_\perp$ drives a transition from G-type AF to 
A-type AF($0,0,\pi$) order, where the critical $t'_\perp$ value remains nearly constant.
Adding moderate in-plane frustration with $t'_{xy} = 0.5$ [see Fig.~\ref{fig11}(c)]
preserves the low-$t_z$ behavior but modifies the high-$t_z$ regime. Here, the
G-type AF phase first transitions into the doubly degenerate C-type
AF[($0,\pi,\pi$)/($\pi,0,\pi$)] state, before transforming to the A-type
AF($0,0,\pi$) phase at larger $t'_\perp$, covering almost the whole tensile regime.
To complete our ground state analysis of NNN hopping anisotropy, we present the
results for large in-plane NNN hopping ($t'_{xy}=1$) in Fig.~\ref{fig11}(d). Here,
the doubly degenerate C-type AF[($0,\pi,\pi$)/($\pi,0,\pi$)] phase dominates a
large portion of the phase diagram compared to the $t'_{xy}=0.5$ case,
effectively suppressing the entire G-type AF ordering region and part of the
A-type AF($0,0,\pi$) phase.

\begin{figure}[t]
\centering
\includegraphics[width=0.48\textwidth]{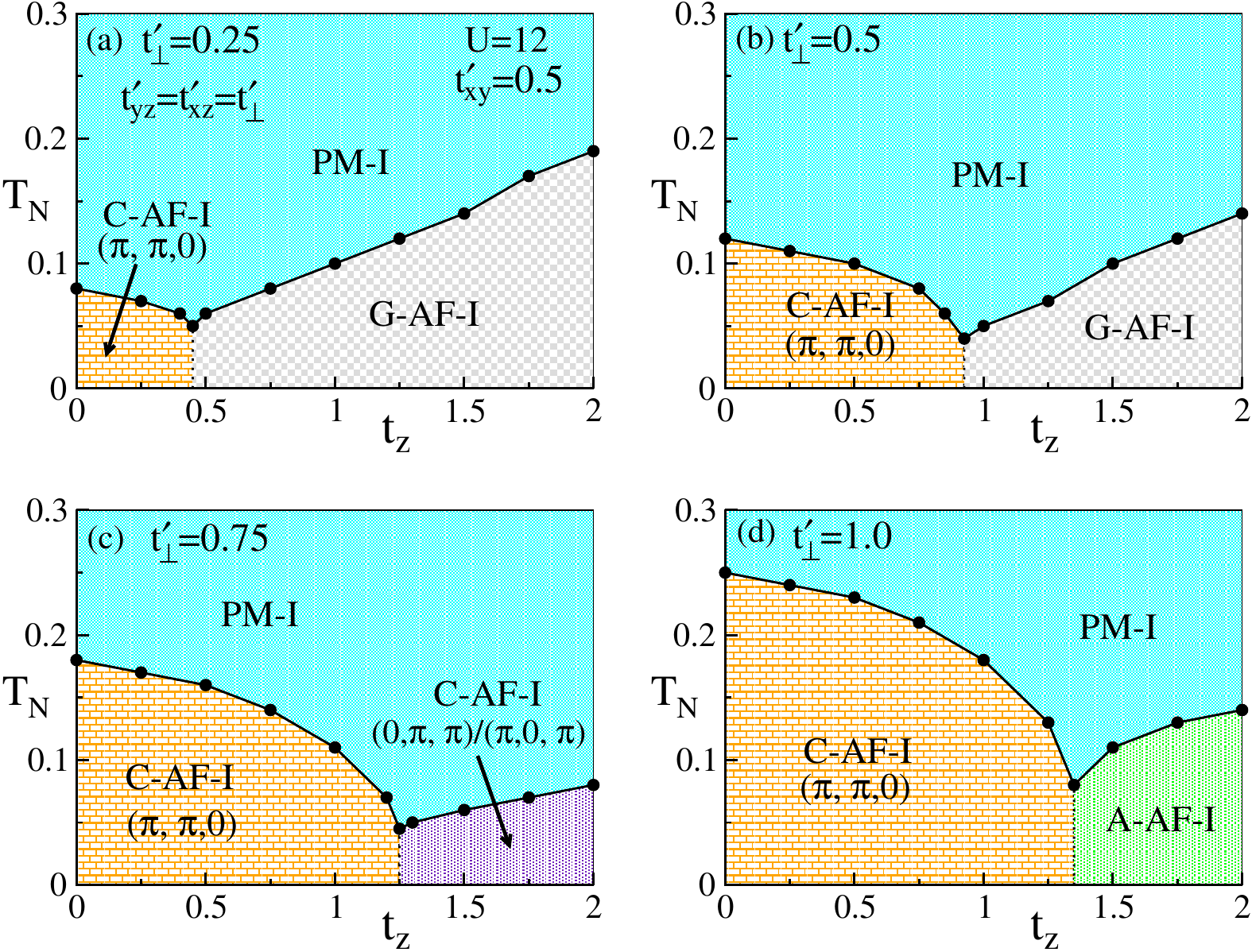}
\caption{The $t_z$-$T_N$ phase diagrams for fixed in-plane NNN hopping
$t'_{xy} = 0.5$ at various out-of-plane hoppings: (a) $t'_{\perp} = 0.25$,
(b) $t'_{\perp} = 0.5$, (c) $t'_{\perp} = 0.75$, and (d) $t'_{\perp} = 1.0$.
For small-to-moderate $t'_{\perp}$ ($0.25$ and $0.5$), a C-type AF($\pi,\pi,0$)
phase is stabilized at low $t_z$; its $T_N$ decreases as $t_z$ increases until
a transition to the G-type AF phase occurs near $t_z \sim 2 t'_{\perp}$.
At $t'_{\perp} = 0.75$, the G-type AF phase is suppressed at large $t_z$,
giving way to a twofold-degenerate C-type AF[$(0,\pi,\pi)/(\pi,0,\pi)$] phase for
$t_z \gtrsim 1.25$. In the large-hopping regime ($t'_{\perp} = 1.0$), an A-type 
AF($0,0,\pi$) phase emerges beyond $t_z \approx 1.35$. In panels (c) and (d), 
$T_N$ also decreases with $t_z$ within the initial phase and then increases
after the magnetic transition into the subsequent phase.
}
\label{fig12}
\end{figure}

Finally, we present the evolution of the $T_N$ as a function of the out-of-plane NN
hopping $t_z$ for different out-of-plane NNN hoppings $t'_\perp$, while fixing
$t'_{xy} = 0.5$. This analysis complements the low-temperature phase diagram
discussed in Fig.~\ref{fig11}(c), which comprises all possible phases. For
small and intermediate $t'_\perp$ values [Figs.~\ref{fig12}(a) and (b)], the
$T_N$ of the C-type AF($\pi,\pi,0$) phase decreases with increasing $t_z$
until it transitions to a G-type AF phase around $t_z \sim 2t'_\perp$. Upon
further enhancing $t'_\perp$ to $0.75$, the G-type AF phase is no longer
stable at high $t_z$ values; instead, the twofold-degenerate C-type
AF[($(0,\pi,\pi)$/$(\pi,0,\pi)$)] phase becomes stabilized (beyond $t_z \approx 1.25$),
as shown in Fig.~\ref{fig12}(c). For a strong NNN hopping, $t'_\perp = 1$
[see Fig.~\ref{fig12}(d)], the magnetic behavior is qualitatively distinct: the
PM-I state evolves into the C-type AF($\pi,\pi,0$) phase below $t_z \sim 1.35$,
but an A-type AF($0,0,\pi$) phase emerges above this threshold. Notably, in all
cases, the $T_N$ decreases monotonically with $t_z$ until the magnetic transition,
after which it increases. Crucially, the $T_{\text{MIT}}$ consistently surpasses
$T_N$, which ensures the existence of a stable PM-I regime at finite temperatures
above the magnetically ordered phases.

\section{Conclusions}\label{sec_con}

We investigated the control of magnetic properties and competing AF phases in
correlated electron systems by taking $\rm BiFeO_3$ as model system using epitaxial
strain as the tuning parameter. By combining first-principles DFT calculations with s-MC
simulations on an anisotropic Hubbard model, we demonstrated a direct microscopic link
between strain-induced structural distortion and magnetic phase transitions.
Our DFT analysis first identified a direct microscopic link: it revealed that
the compressive epitaxial strain induces a structural distortion in $\rm BiFeO_3$
that drives a transition from the G-type AF phase to a C-type AF($\pi,\pi,0$) phase
by introducing significant anisotropy in exchange interactions. 
Subsequently, the s-MC phase diagrams, which simulate the effect of strain through
anisotropic hopping parameters ($t_z$ and $t'_\perp$) informed by the DFT analysis,
corroborate this G-type to C-type crossover and demonstrate that the in-plane NNN
hopping $t'_{xy}$ plays no dominant role in driving this transition. Furthermore,
the s-MC calculations predict that tensile strain reorganizes the AF phase landscape,
ultimately stabilizing an A-type AF($0,0,\pi$) phase as well as a twofold-degenerate
C-type AF[($(0,\pi,\pi)$/$(\pi,0,\pi)$)] phase.  
This integrated study thus established a mechanism where epitaxial strain,
by governing directional hopping anisotropies, acts as a potent parameter for
engineering competing AF states, leading to a near-degeneracy of multiple AF phases
that holds promise for tunable magnetic phase transitions in magnetoelectric
and spintronic applications.


\section*{acknowledgment}
We acknowledge use of the Meghnad2019 computer cluster at SINP.


\appendix

\section{Derivation of the Effective Spin-Fermion Hamiltonian} \label{derivation_Heff}
To implement the s-MC framework, we begin by reformulating the quartic on-site 
interaction term in the Hubbard Hamiltonian [in Eq.~\ref{hamiltonian}] using the 
Hubbard--Stratonovich transformation. This decoupling introduces auxiliary fields 
that facilitate a tractable quadratic form. First, we express the on-site interaction 
term as:
\begin{equation} \label{dh_1}
n_{i,\uparrow} n_{i,\downarrow} = \frac{1}{4} n_i^2 - S_{iz}^2 = \frac{1}{4} n_i^2 - \left( \mathbf{S}_i \cdot \hat{\Omega}_i \right)^2,
\end{equation}
where $\mathbf{S}_i = \frac{1}{2} \sum_{\alpha\beta} c_{i,\alpha}^\dagger \boldsymbol{\sigma}_{\alpha\beta} c_{i,\beta}$ 
is the spin operator at site $i$ (with $\hbar = 1$), and $\boldsymbol{\sigma}$ denotes 
the Pauli matrices. The arbitrary unit vector $\hat{\Omega}_i$ appears due to 
SU(2) symmetry, allowing rotational invariance: 
$S_{ix}^2 = S_{iy}^2 = S_{iz}^2 = (\mathbf{S}_i \cdot \hat{\Omega}_i)^2$ 
for any $\hat{\Omega}_i$.

The partition function for the Hamiltonian stated in Eq.~\ref{hamiltonian} 
is defined as $Z = \mathrm{Tr}[e^{-\beta H}]$, 
with the trace taken over the all particle numbers and site occupations. 
The inverse temperature is $\beta = 1/T$, and we set $k_B = 1$. 
To evaluate $Z$ numerically, we discretize the imaginary time interval $[0, \beta]$ 
into $M$ steps of width $\Delta\tau$ such that $\beta = M\Delta\tau$, 
and apply a first-order Suzuki--Trotter decomposition:
\begin{equation}
e^{-\beta(H_0 + H_I)} \approx \left( e^{-\Delta \tau H_0} e^{-\Delta \tau H_I} \right)^M + \mathcal{O}(\Delta\tau^2).
\end{equation}

Using Eq.~\ref{dh_1}, the interaction term at a single time slice $l$ can be 
decoupled into a path integral over scalar and vector auxiliary fields:
\begin{align}
&\int d\phi_i(l) \, d\Delta_i(l) \, d^2\Omega_i(l) \, \exp\Bigg[-\Delta\tau \sum_i \Bigg\{ \frac{\phi_i^2(l)}{U} + i \phi_i(l) n_i \nonumber \\
&\qquad + \frac{\Delta_i^2(l)}{U} - 2\Delta_i(l) \hat{\Omega}_i(l) \cdot \mathbf{S}_i \Bigg\} \Bigg],
\end{align}
where $\phi_i(l)$ and $\Delta_i(l)$ are the auxiliary fields for charge and 
spin density, respectively. We define a vector auxiliary field 
$\mathbf{m}_i(l) = \Delta_i(l) \hat{\Omega}_i(l)$ to simplify notation. 
The full partition function becomes:
\begin{align}
Z &\propto \mathrm{Tr} \prod_{l=M}^1 \int d\phi_i(l) \; d^3\mathbf{m}_i(l) \exp \Bigg[-\Delta\tau \Bigg\{ H_0 + \sum_i \nonumber \\
& \Bigg( \frac{\phi_i^2(l)}{U} + i \phi_i(l) n_i + \frac{\mathbf{m}_i^2(l)}{U} - 2 \mathbf{m}_i(l) \cdot \mathbf{S}_i \Bigg) \Bigg\} \Bigg].
\end{align}
At this stage, the formulation is exact, preserving full SU(2) spin symmetry, 
with auxiliary fields depending on both space and imaginary time.

To make the problem computationally feasible within the s-MC approach, 
we adopt two approximations: (i) we drop the imaginary-time dependence 
of the auxiliary fields, treating $\phi_i$ and $\mathbf{m}_i$ as classical fields; 
and (ii) we fix the charge field to its saddle-point value, 
$i\phi_i = \frac{U}{2} \langle n_i \rangle$. With a rescaling 
$\mathbf{m}_i \rightarrow \frac{U}{2} \mathbf{m}_i$, 
we obtain the following effective Hamiltonian:
\begin{align}
H_\text{eff} = & -t \!\! \sum_{\left\langle i,j \right\rangle , \sigma} \!\! c_{i,\sigma}^\dagger c_{j,\sigma} -t' \!\!\!\! \sum_{\left\langle \left\langle i,j \right\rangle \right\rangle, \sigma} \!\!\!\! c_{i,\sigma}^\dagger c_{j,\sigma} \nonumber \\ 
 & + \frac{U}{2} \sum_i \left( \left\langle n_i \right\rangle n_i - \mathbf{m}_i \cdot {\boldsymbol{\sigma}}_i \right) \nonumber \\ 
 &+ \frac{U}{4} \sum_i \left( \mathbf{m}_i^2 - \left\langle n_i \right\rangle^2 \right) - \mu \sum_i n_i,
\end{align}
as presented earlier in Eq.~\ref{h_eff}. The resulting spin-fermion model describes 
itinerant electrons interacting with classical auxiliary fields $\{ \mathbf{m}_i \}$, 
which are sampled using Monte Carlo techniques.

\section{Definitions of Physical Observables} \label{obs}

To characterize the magnetic and transport properties of the system, 
we compute several observables, including the magnetic structure factor, 
local magnetic moment, specific heat, density of states, and resistivity. 
These quantities serve as key indicators of phase transitions and ordering 
tendencies within the spin-fermion model governed by Eq.~\ref{h_eff}.

\textbf{Magnetic correlations:}
The magnetic ordering is quantified using the spin structure factor $S(\mathbf{q})$, 
which captures long-range spin correlations. It is defined as
\begin{align}
S(\mathbf{q}) = \frac{1}{(L^3)^2} \sum_{k,l} \left\langle \mathbf{S}_k \cdot \mathbf{S}_l \right\rangle e^{-i \mathbf{q} \cdot (\mathbf{r}_k - \mathbf{r}_l)},
\end{align}
which is measured in units of $\hbar^2$.
$\mathbf{S}_k$ denotes the spin operator at site $k$, and the angular brackets 
indicate combined quantum and thermal averages. The sum runs over all pairs of 
lattice sites on a system of volume $L^3$, where $L$ is the length of the sides 
of the simple cubic lattice. Specific wave vectors $\mathbf{q}$ correspond to 
distinct magnetic configurations (for detail see Fig.~\ref{fig01}): 
$\mathbf{q} = (\pi,\pi,\pi)$ indicates G-type AF phase; 
$\mathbf{q} = (\pi,\pi,0), (0,\pi,\pi), (\pi,0,\pi)$ denote C-type AF orders; 
while $\mathbf{q} = (0,0,\pi), (\pi,0,0), (0,\pi,0)$ signal A-type AF alignment.

\textbf{Local moment:}
The local magnetic moment $M$ is defined as the average of the square of quantum magnetization:
\begin{align}
M = \left\langle (n_{\uparrow} - n_{\downarrow})^2 \right\rangle = \left\langle n \right\rangle - 2 \left\langle n_{\uparrow} n_{\downarrow} \right\rangle,
\end{align}
with units of $\hbar^2$.
$\left\langle n \right\rangle = \left\langle n_{\uparrow} + n_{\downarrow} \right\rangle$ 
is the average electron density of the system. At half-filling ($\langle n \rangle = 1$), 
the moment is bounded between $M = 0.5$ (uncorrelated or high-temperature limit) 
and $M = 1$ (large coupling $U \to \infty$, where double occupancy vanishes). 
Thus, the evolution of $M$ with temperature and interaction strength provides 
insight into local moment formation.

\textbf{Specific heat:}
The specific heat $C_v$ is computed as the derivative of the total energy with 
respect to temperature (central difference method is used for the derivative):
\begin{align}
C_v = \frac{dE}{dT}.
\end{align}
Here, $E$ includes contributions from both fermionic degrees of freedom 
(quantum energy) and the classical auxiliary fields (classical energy). 
Numerically, we evaluate $C_v$ using central difference scheme.

\textbf{Density of states:}
To probe the electronic spectrum, we calculate the DOS, defined as
\begin{align}
N(\omega) = \sum_{i=1}^{2L^3} \delta(\omega - \omega_i),
\end{align}
where $\omega_i$ are the eigenvalues of the effective Hamiltonian in Eq.~\ref{h_eff}. 
In practice, the $\delta$-function is replaced by a Lorentzian with a broadening of $\sim W/(2L^3)$, 
where $W$ denotes the bandwidth of the noninteracting model. 
The factor $2L^3$ corresponds to the total number of single-particle states.
The Fermi energy is taken to be $\omega = 0$ in all spectral analyses.

\textbf{Resistivity:}
Transport properties are evaluated via the $dc$ limit of the optical conductivity, 
which is computed using the Kubo-Greenwood formula~\cite{Mahan, Kumar}:
\begin{align}
\sigma_z(\omega) = \frac{\pi e^2}{L^3 \hbar a_0} \sum_{\alpha,\beta} (n_\alpha - n_\beta) \frac{|f^z_{\alpha\beta}|^2}{\omega_\beta - \omega_\alpha} \delta\left(\omega - (\omega_\beta - \omega_\alpha)\right),
\end{align}
where $a_0$ is the lattice parameter and we have expressed $\sigma_z$ 
in units of $\frac{e^2}{\hbar a_0}$. $n_\alpha = f(\mu - \omega_\alpha)$ 
is the Fermi-Dirac distribution and 
$f^z_{\alpha\beta} = \langle \psi_\alpha | J_z | \psi_\beta \rangle$ 
denotes the current matrix element along the $z$ direction. 
The current operator along the $z$ direction is defined as 
\begin{align*}
J_z = i a_0 \sum_{i,\sigma} &t_z (c_{i,\sigma}^\dagger c_{i+a_0 \hat{z},\sigma} - \text{H.c.}) \\ &+t'_{xz} ( c_{i,\sigma}^\dag c_{i+a_0\hat{z}+a_{0}\hat{x},\sigma} - \text{H.c.}) \\ &+t'_{yz} ( c_{i,\sigma}^\dag c_{i+a_0\hat{z}+a_{0}\hat{y},\sigma} - \text{H.c.}).
\end{align*}
The eigenvalues $\omega_\alpha$ and corresponding eigenstates $\psi_\alpha$ 
are obtained from diagonalizing $H_{\text{eff}}$. The $dc$ conductivity 
(along the $z$-axis) $\sigma^z_{\text{dc}}$ is estimated by integrating 
$\sigma_z(\omega)$ over a small low-frequency interval:
\begin{align}
\sigma^z_{\text{dc}} = \frac{1}{\Delta\omega} \int_0^{\Delta\omega} \sigma_z(\omega) \, d\omega.
\end{align}
The parameter $\Delta \omega$ is chosen to be approximately four to five times 
larger than the system’s average finite-size gap, 
which is estimated as the total bandwidth divided by the number of eigenvalues.
The resistivity along the $z$ direction is defined as $\rho_z = 1/\sigma^z_{\text{dc}}$.



\end{document}